\documentclass[draft]{agujournal2019}
\usepackage{url}
\usepackage{lineno}
\usepackage{amssymb}
\usepackage{amsmath}
\usepackage{soul}
\usepackage{xfrac}
\usepackage{multirow,tabularx}
\usepackage{ragged2e}
\justifying

\usepackage{xcolor}

\newcommand{\OL}{\overline}

\newcommand{\btimes}{{\mbox{\boldmath $\times$}}}

\newcommand{\be}{\begin{equation}}
\newcommand{\ee}{\end{equation}} 
\newcommand{\lb}{\label}

\newcommand{\ME}{{\mathcal{E}}}
\newcommand{\olE}{{\overline{E}}}

\newcommand{\paren}[1]{\left( #1 \right)}

\newcommand{\bracep}[1]{\left\{ #1 \right\}}
\newcommand{\absv}[1]{\left| #1 \right|}
\newcommand{\angp}[1]{\left< #1 \right>}

\newcommand{\br}{{\bf r}}
\newcommand{\bu}{{\bf u}}

\newcommand{\bx}{{\bf x}}

\newcommand{\phz}{\hphantom{0}}
\newcommand{\phpz}{\hphantom{.00}}
\draftfalse

\journalname{JAMES Journal of Advances in Modeling Earth Systems}

\begin{document}

%% ------------------------------------------------------------------------ %%
%  Title
%
% (A title should be specific, informative, and brief. Use
% abbreviations only if they are defined in the abstract. Titles that
% start with general keywords then specific terms are optimized in
% searches)
%
%% ------------------------------------------------------------------------ %%

\title{Spatio-temporal coarse-graining decomposition of the global ocean geostrophic kinetic energy}

%% ------------------------------------------------------------------------ %%
%
%  AUTHORS AND AFFILIATIONS
%
%% ------------------------------------------------------------------------ %%

\authors{M. Buzzicotti\affil{1}, B. A. Storer\affil{2}, H. Khatri\affil{3}, S.M. Griffies\affil{4,5}, and H. Aluie\affil{2,6}}

\affiliation{1}{Department of Physics, University of Rome Tor Vergata and INFN}
\affiliation{2}{Department of Mechanical Engineering, University of Rochester}
\affiliation{3}{Department of Earth, Ocean and Ecological Sciences, University of Liverpool}
\affiliation{4}{NOAA Geophysical Fluid Dynamics Laboratory}
\affiliation{5}{Princeton University Atmospheric and Oceanic Sciences Program}
\affiliation{6}{Laboratory for Laser Energetics University of Rochester}

%% Corresponding Author:
% Corresponding author mailing address and e-mail address:

\correspondingauthor{Hussein Aluie}{hussein@rochester.edu}

\vspace{0.5cm}
\begin{center}
Draft from \today 
\end{center}

\begin{keypoints}

\item Notable hemispheric asymmetry in mesoscale energy-per-area, which is higher in the north due to continental boundaries.

\item Spectra of the time-mean velocity show that most (up to $70\%$) of its energy resides in `standing' mesoscale eddies $<500~$km.

\item We estimate that $\approx25$--50\% of total geostrophic energy is at scales smaller than 100~km, and is un(der)-resolved by pre-SWOT satellite products.

\end{keypoints}

%% ------------------------------------------------------------------------ %%
%
%  ABSTRACT and PLAIN LANGUAGE SUMMARY
%
%
%% ------------------------------------------------------------------------ %%

\begin{abstract} % currently just under the 250 word limit
We expand on a recent determination  of the first global energy spectrum of the ocean's surface geostrophic circulation~\cite{Storer2022} using a coarse-graining (CG) method. We compare spectra from CG to those from spherical harmonics by treating land in a manner consistent with the boundary conditions. While the two methods yield qualitatively consistent domain-averaged results, spherical harmonics spectra are too noisy at gyre-scales ($>1000~$km). More importantly, spherical harmonics are inherently global and cannot provide local information connecting scales with currents geographically. CG shows that the extra-tropics mesoscales (100--500~km) have a root-mean-square (rms) velocity of $\sim15~$cm/s, which increases to $\sim30$--40~cm/s locally in the Gulf Stream and Kuroshio and to $\sim16$--28~cm/s in the ACC. There is notable hemispheric asymmetry in  mesoscale energy-per-area, which is higher in the north due to continental boundaries. 
We estimate that $\approx25$--50\% of total geostrophic energy is at scales smaller than 100~km, and is un(der)-resolved by pre-SWOT satellite products.
Spectra of the time-mean component show that most of its energy (up to $70\%$) resides in stationary mesoscales ($<500~$km), highlighting the preponderance of `standing' small-scale structures in the global ocean. 
By coarse-graining in space and time, we compute the first spatio-temporal global spectrum of geostrophic circulation from AVISO and NEMO. These spectra show that every length-scale evolves over a wide range of time-scales with a consistent peak at $\approx200$ km and $\approx2$--3~weeks. 
\end{abstract}

\section*{Plain Language Summary} % currently at ~155 words  (200 word limit)
Traditionally, `eddies' are identified as time-varying features relative to a background time-mean flow. As such, `mean' does not imply large length-scale. Standing eddies or meanders due to topography have little time-variation, but can have significant energy at small length-scales that are unresolved and need to be parameterized in coarse climate simulations. Similarly, `eddy' or `time-varying' do not imply small length-scale, such as large-scale motions from Rossby waves or fluctuations of the Kuroshio. Another common method is Fourier analysis in `representative' ocean boxes that cannot capture the circulation’s planetary scales. We overcome these limitations thanks to recent advances: (i) a method for calculating spectra by coarse-graining, (ii) properly defining convolutions on the sphere, which `blur' oceanic flow in a way that preserves its underlying symmetries, opening the door for global `wavelet’ analysis and, more generally, spatial coarse-graining, and (iii) FlowSieve: an efficient parallel code. We employ coarse-graining in space-time to gain new insights into the global oceanic circulation, including how much energy resides in its different spatial structures and how they vary in time. 

%% ------------------------------------------------------------------------ %%
%
%  TEXT
%
%% ------------------------------------------------------------------------ %%

\section{Introduction}
\label{section:intro}

Ocean circulation emerges from a suite of linear and nonlinear dynamical processes that act  over a broad range of spatial and temporal scales. The flow field is markedly inhomogeneous and characterized by waves, instabilities, and turbulent eddies, each of which are subject to a variety of energetic sources and sinks.
The mesoscale defines a key band of spatial scales where ocean flows are largely geostrophic and where kinetic energy peaks \cite{wunsch2007past,Storer2022}. Correspondingly, it is widely recognized that flow at the ocean mesoscales, and its response to changes in atmospheric forcing, are fundamental to the large-scale circulation and central for regional and global transport of heat and biogeochemical tracers \cite{FerrariWunsch09}. 

However, significant gaps remain in our understanding of the mesoscale flows and their role in ocean circulation and climate. In particular, from a numerical modeling perspective, despite the ever-increasing ability to conduct simulations with mesoscale eddy-rich OGCM, accurately resolving these scales in routine climate-scale (order centuries and longer) simulations remains the exception rather than the norm \cite<e.g. see>{Griffiesetal15}. We are thus confronted with the need for mesoscale eddy  parameterizations for the foreseeable future \cite{Pearsonetal17}.

A central question of physical oceanography, and in particular the eddy parameterization problem, concerns a characterization of flow features according to length-scale. This question motivates the goal of this paper, which is to provide a length-scale decomposition of the global ocean geostrophic kinetic energy, and to study the seasonal variations of this decomposition. This  goal has previously been out of reach due to limitations of the commonly used Fourier spectral methods, which are unsuited to global ocean analysis due to the complex geometry of ocean basins. We thus make use of a Coarse-Graining (CG) method that does not share the limitations of Fourier analysis. This paper  serves to detail the use of coarse-graining for the purpose of decomposing ocean kinetic energy, and in so doing we uncover novel features of the ocean surface circulation as a function of length and time scales.

\subsection{Fourier methods for the ocean}
\label{sec:Fourier}

It is common to quantify the spectral distribution of ocean kinetic energy via Fourier transforms computed either along transects or within regions; e.g.,  \citeA{FuSmith1996,Chenetal2015,Rochaetal2016,khatri2018surface, ORourke_etal2018, CalliesWu2019}. 
This approach has rendered great insights into the length scales of oceanic motion and the cascade of energy through these scales \cite{ScottWang05,ScottArbic07,arbic2012nonlinear,Arbicetal13,arbic2014geostrophic}. 
However, it has notable limitations for the ocean where the spatial domain is generally not periodic, thus necessitating adjustments to the data (e.g., by tapering) before applying Fourier transforms. 

Methods to produce an artificially periodic dataset can introduce spurious gradients, length-scales, and flow features not present in the original data \cite{sadek2018extracting}. 
A related limitation concerns the chosen region size, with this size introducing an artificial upper length scale cutoff. 
In this manner, no scales are included that are larger than the region size even if larger structures exist in the ocean. 
Furthermore, the data is typically assumed to lie on a flat tangent plane to enable the use of Cartesian coordinates. 
However, if the region becomes large enough to sample the earth's curvature, then that puts into question the use of the familiar Cartesian Fourier analysis of sines and cosines. 

We have previously compared coarse-graining methods with traditional Fourier methods, and shown that where Fourier methods are valid, both methods agree \cite{Storer2022}. An important advantage of coarse-graining is that it is not limited to an ocean box and allows us to probe length-scales extending to the planet's circumference. Moreover, unlike Fourier analysis in box regions, which cannot account for the global energy in the ocean, coarse-graining satisfies energy conservation \cite{sadek2018extracting} as we discuss more below.

Spherical harmonics transform is another Fourier (or spectral) method over the entire globe, often used in atmospheric modeling \cite{satoh2004atmospheric}. It is seldom applied to the ocean due to continental boundaries. Spherical harmonics are basis functions that are defined over the entire sphere and are not restricted to the ocean domain. During early days of satellite altimetry, there were attempts at utilising the method to characterize the frequency-wavenumber spectrum of the ocean's global circulation \cite{wunsch1991global,wunsch1995global}. These studies analyzed sea surface height (SSH) anomalies and chose nominal SSH values over land. SSH over land was set to the time average of the zonal mean absolute topography. However, the authors were aware that their choice for land treatment was somewhat \textit{ad hoc}, without dynamical justification, as stated in \citeA{wunsch1995global}: ``\dots we make no claim that we have made the best possible choice.'' It seems that usage of spherical harmonics was largely abandoned after these attempts during the early days of satellite altimetry.
In this paper, we revisit spherical harmonics transform in section~\ref{sec:spherical_harmonics} and show that despite its important limitations, the method can yield meaningful results if land is treated in a manner that is consistent with boundary conditions of the ocean's dynamics.

\subsection{Eddy and mean flow decomposition: Reynolds averages}

A traditional approach to extract `eddies' from a flow uses time or ensemble averaging. 
This approach is relatively simple operationally and is in accord with the common practice in atmospheric and oceanic sciences of studying long-term climate means and fluctuations relative to that mean. 
As part of this decomposition for turbulent flow, we typically utilize the time averaging operator as a Reynolds averaging (RA) operator, whereby the average of a fluctuating quantity vanishes \cite{Vallis17}. 
The choice of Reynolds decomposition by time averaging is largely based on practical considerations, with ensemble averages being unavailable for most applications (although see \cite{Uchidaetal2021} for a recent example with fine resolution regional ocean simulations). 
 
Within the traditional decomposition, time-mean or ensemble-mean do not necessarily imply a large length-scale flow as we shall discuss in this paper. 
For example, standing eddies or stationary meanders due to topography \cite{Youngsetal2017} have little temporal or statistical fluctuations but can have spatial structure at length-scales $\mathcal{O}(100)~$km or smaller. 
Similarly, within a Reynolds decomposition, `eddy' does not necessarily imply small length-scale.  
For example, a time averaging based decomposition would ascribe eddying motion to large-scale Rossby waves \cite{kessler1990observations} or variations in the Kuroshio Current's path \cite{kawabe1995variations}. 
 
By construction, a Reynolds decomposition into a mean and an `eddy' limits our ability to analyze temporal variability, from intra-annual to inter-annual \cite{Bryanetal14,Griffiesetal15}, of the multiscale coupling and evolution of different length-scales, including those that need to be resolved/predicted in global climate (coarse-grid) models. 
Therefore, it offers limited guidance for coarse-resolution models and no control over the specific physical length which partitions oceanic flow into `large' and `small'. 
In other words, the set of length-scales constituting the large-scale flow cannot be varied/controlled to be consistent with those length-scales resolved in a coarse climate simulation. 
In this sense, the traditional mean-eddy decomposition cannot help with on-going efforts to develop `scale-aware' parameterizations \cite{Ringleretal13,Zannaetal17,Pearsonetal17,jansen2019toward}, including those using data-driven or machine learning approaches \cite{ryzhov2020data,ross2023benchmarking}. 

\subsection{Empirical Orthogonal Functions}
Empirical Orthogonal Functions (EOFs) offer yet another approach for decomposing the oceanic flow by projecting onto orthogonal basis functions or `empirical modes' that are derived from the data itself. 
EOF is also known as Karhunen-Loeve decomposition, Principal Component Analysis (PCA) or Proper Orthogonal Decomposition (POD) in other fields \cite{KacSiegert1947,Karhunen1947,Loeve1948}, and was introduced to meteorology by \citeA{Lorenz1956}. 
 
EOF analysis is commonly used as a data reduction technique since it offers the most efficient statistical compression of the data field \cite{ThomsonEmery01}. 
This is because the basis functions are derived from the statistical analysis of the data and do not necessarily correspond to true dynamical modes, although they have yielded valuable insight into the oceanic dynamics on climate scales \cite<e.g.>{Trenberth1975,DiLorenzo2008}. 
The limitation of EOFs stems from our lack in understanding of the dynamics governing the basis functions. 
Moreover, it is difficult to associate EOFs with lengthscales or timescales since each empirical mode lumps together variations over all frequency and wavenumber bands. 
This approach muddles the interpretation of EOF spectra and their connection to spectral slopes predicted by theory \cite{Uchidaetal2021}.

\subsection{Coarse-graining}
In order to understand the multiscale nature of oceanic flows, while simultaneously resolving them in space and in time, we use a `coarse-graining' framework that is relatively new in physical oceanography \cite{aluie2018mapping,BuseckeAbernathey2019,srinivasan2019submesoscale,schubert2020submesoscale,Raietal2021,barkan2021oceanic,haigh2021eddy,khani2023gradient,loose2023diagnosing}.
It is a very general approach to decompose complex flows, with rigorous foundations initially developed to model \cite{Germano92,Meneveau1994} and analyze \cite{Eyink95,Eyink05} turbulence. 
\citeA{Aluie17} provides a theoretical discussion of coarse-graining and its connection to other methods in physics. 
The approach has been recently generalized to account for the spherical geometry of flow on Earth \cite{aluie2019convolutions}, and applied to study the nonlinear cascade in the North Atlantic from an eddying simulation \cite{aluie2018mapping}. 

The coarse-graining framework is very useful from the standpoint of ocean subgrid scale parameterizations \cite{fox2011parameterization,Zannaetal17,khani2019diagnosing,jansen2019toward,haigh2020tracer,stanley2020vertical,grooms_etal2021}. 
Namely, it provides a theoretical basis for constructing subgrid closures that faithfully reflect the dynamics at unresolved scales.
A primary objective in ocean modeling is practical: an accurate subgrid parameterization that is numerically stable. 
Significant advances have been achieved in this regard in the fluid dynamics and turbulence community \cite{Piomellietal91,buzzicotti2018energy,linkmann2018multi,Biferale2019SelfSim,di2020phase,buzzicotti2020synch}, and the field of large-eddy simulation (LES) is well-established \cite{meneveau2000scale}. 

Our use of coarse-graining supports the needs of parameterization, but our primary objective is to characterize the fundamental dynamics of the flow at \emph{all} length scales. 
Even within the wider fluid dynamics community, much less work has been done in this regard, i.e. using coarse-graining as a `probe' of the fundamental scale-physics.  
For example, LES sub-grid parameterization studies are seldom concerned with using coarse-graining to probe the energy pathways across the entire range of scales, such as the cascade \cite{Eyink95,EyinkAluie09,KelleyOuellette11,Aluieetal12,Riveraetal14,buzzicotti2018effect,buzzicotti2021inertial}, forcing \cite{Aluie13,Raietal2021}, dissipation \cite{ZhaoAluie18}, or the range of coupling between different scales \cite{Eyink05,AluieEyink09}.

As an important case in point, despite LES being a well established field in fluid dynamics since the seminal works of \citeA{Leonard74} and \citeA{germano1992turbulence}, the idea of using coarse-graining in physical space to extract the energy content at different scales; i.e.,  the spectrum, was only recently established and demonstrated by \citeA{sadek2018extracting}. 
This method is central to our calculation here of the spectrum for the oceanic general circulation. 
A main advantage of coarse-graining is that it allows us to decompose different length scales in a flow, at any geographic location and any instant of time, without relying on assumptions of homogeneity, isotropy or domain periodicity. 
This generality makes it ideally suited for studying oceanic flows with complex continental boundaries over the entire globe or in any particular regions of interest and at any time.

\subsection{Key results and outline of this paper}

In this paper we make use of the coarse-graining method on a satellite sea surface product and an Ocean General Circulation Model (OGCM) simulation. To directly compare the two products, we focus on geostrophic components of the horizontal surface velocity as diagnosed from sea level. 
Here, we highlight key novel results from this analysis. First, we show that spectra from coarse-graining and spherical harmonics of the global circulation are consistent but the latter cannot yield spatially local information. We show that the typical velocity of mesoscales is of the order of $15$~cm/s, but reaches 30--40~cm/s in western boundary currents (WBCs) and 16--28~cm/s in the ACC. We find notable hemispheric asymmetry in mesoscale energy-per-area, which is higher in the north. This asymmtery is compensated by the south having more energy-per-area at gyre-scales, such that across all (resolved) scales, the two hemispheres have comparable energy-per-area.
From our spectra, we can estimate that  $\approx25$--$50\%$ of total geostrophic energy is at scales smaller than 100~km, and is un(der)-resolved by pre-SWOT satellite products. Spectra of the time-mean velocity show that most (up to $70\%$) energy resides in `standing' mesoscale eddies $<500~$km. By coarse-graining in space and time, we compute the first spatio-temporal global spectrum of geostrophic circulation from AVISO and NEMO. These spectra show that every length-scale evolves over a wide range of time-scales with a consistent peak at $\approx200$ km and $\approx3$ weeks.

The paper is organized as follows. In Section~\ref{sect:Datasets}, we present the data products used in our analysis. In Section~\ref{sect:methods} we give details on the coarse-graining and the Reynolds averaging methods used in this work and we present the comparison between CG and spherical harmonics energy spectra. In Section~\ref{sect:Results} we discuss the main results from the CG analysis; the 2D spatio-temporal energy spectrum of ocean surface circulation and spectra of the time-mean and fluctuating (or `eddy') components from Reynolds averaging. At the end of Section~\ref{sect:Results} we compare the surface dynamics spatio-temporal decomposition from satellite and numerical model data. In Section~\ref{sect:Conclusions} we present our conclusions. \ref{appendix:deforming_kernel} discusses some technical choices we used when coarse-graining.

\section{Satellite and numerical model data}
\label{sect:Datasets}

We examine the horizontal geostrophic velocity of surface ocean currents from a global  numerical model simulation and from an analysis of satellite sea surface altimetry, focusing on regions to the north and south of the tropics, $[15^\circ \text{N} - 90^\circ \text{N}]$ and $[15^\circ \text{S} - 90^\circ \text{S}]$.  
We avoid the tropics since our interest is with the geostrophic flows in the higher latitudes, and only the surface geostrophic current is available from satellite altimetry.
Details of the two products are given in the following paragraphs, and both were publicly accessed through the Copernicus Marine Environment Monitoring Service (CMEMS) webpage, \url{https://marine.copernicus.eu/services-portfolio/access-to-products/}. 

\paragraph*{AVISO analysis of satellite altimetry}

Geostrophic currents are obtained from the AVISO$+$ analysis of multi-mission satellite altimetry measurements for sea surface height (SSH) \cite{pujol2016}. We used the Level 4 (L4) post-processed dataset of daily-averaged geostrophic velocity, gridded at a resolution of ${0.25^\circ \times 0.25^\circ}$ and spanning from January 2010 to October 2018. Post processing was performed by the Sea Level Thematic Center (SL TAC) data processing system, which processes data from eleven altimeter missions.
The product identifier of the AVISO dataset used in this work is ``\sloppy{{\small SEALEVEL\_GLO\_PHY\_L4\_MY\_008\_047}}'' {(\url{https://doi.org/10.48670/moi-00148})}.

\paragraph*{Numerical simulation}

We  analyze 1-day averaged surface geostrophic currents from the NEMO numerical modeling framework, which is coupled to the Met Office Unified Model atmosphere component, and the Los Alamos sea ice model (CICE). 
The NEMO dataset consists of weakly coupled ocean-atmosphere data assimilation and forecast system, with data then published on a uniform $\sfrac{1}{12}^\circ$ grid. 
We use daily-averaged data that spans the four years from 2016 to 2019.  
More details about the coupled data assimilation system used for the production of the NEMO dataset can be found in~\cite{gmd-4-223-2011,lea2015assessing}. 
The specific product identifier of the NEMO dataset used here is ``\sloppy{{\small GLOBAL\_MULTIYEAR\_PHY\_001\_030}}'' {(\url{https://doi.org/10.48670/moi-00021})}.

\section{Coarse-graining for the ocean}
\label{sect:methods}

In this section, we discuss the coarse-graining framework and how it is used to partition energy across length scales. We also discuss the traditional approach of decomposition in spherical harmonics and the temporal-based  Reynolds averaging, in which the flow is decomposed into time-mean and fluctuating components.

\subsection{Basics of coarse-graining on the sphere}
\label{sect:meth-CG}

For any scalar field, $F(\bx)$, we can calculate its coarse-grained (or low-pass filtered) version, $\OL F_\ell(\bx)$, by convolving $F(\bx)$ with a normalized filter kernel $G_\ell(\br)$, 
\begin{linenomath*}
\be
\OL F_\ell(\bx) = G_\ell * F(\bx)
\lb{def:filtering}
\ee
\end{linenomath*}
where $*$, in the context of this work, is convolution on the sphere \cite{aluie2019convolutions}, $\bx$ is geographic location on the globe, and the kernel $G_\ell(\bx)$ can be any non-negative function that is spatially localized (\textit{i.e.} it goes to zero fairly rapidly as ${\bx\to\pm\infty}$). The parameter $\ell$ is a length-scale related to the kernel's `width'. We use the notation $\OL{(\cdots)}_\ell$ to denote a  coarse-grained field. 
The kernel is area normalized for all $\ell$, so that
\begin{linenomath*}
\begin{equation}
\int G_\ell(\bx) \, \mathrm{d}\mathcal{S} =1,
\label{eq:normalization-condition}
\end{equation} 
\end{linenomath*}
where $\mathrm{d}\mathcal{S}$ is the area element on the sphere. Correspondingly, the convolution  \eqref{def:filtering} may be interpreted as an average of the function $F$ within a region of diameter $\ell$ centered at location $\bx$. By construction, at each point in space, $\bx$, the coarse-grained field, $\OL{F}_\ell (\bx)$, contains  information about the scale $\ell$.

The above formalism holds for coarse-graining scalar fields. To coarse-grain a vector field on a sphere generally requires more work \cite{aluie2019convolutions}.  However, since we are concerned only with the surface geostrophic velocity, $\bu(\bx,t)$, in this work, it greatly simplifies our analysis. 
We assume the geostrophic velocity is non-divergent on the two-dimensional spherical surface, so that it is related to the geostrophic stream-function $\psi$ via 
\begin{linenomath*}
\begin{equation}
\bu = \hat{\boldsymbol{e}}_r \btimes \nabla \psi,
\lb{eq:geos-vel}\end{equation}
 \end{linenomath*}
with $\hat{\boldsymbol{e}}_r$ the radial unit vector in spherical coordinates, ${\psi = \eta \, g/f}$, $g$ is the gravitational acceleration, $\eta$ the free sea surface height (SSH), and the Coriolis parameter, $f=2\Omega\sin(\phi)$, is a function of latitude $\phi$, where $\Omega$ is Earth's spin rate. 

\begin{figure}
    \centering
    \includegraphics[scale=1]{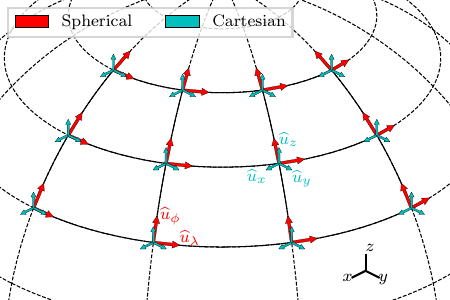}
    \caption{
        Illustration of \textbf{[blue arrows]} Cartesian velocity basis vectors and \textbf{[red arrows]} Spherical velocity basis vectors at selected latitude/longitude points.
        While the spherical basis vectors point in different directions at each location, the Cartesian vectors always point in the same direction.
    }
    \label{fig:VelBasisVectors}
\end{figure}

\citeA{aluie2019convolutions} showed that for non-divergent vector fields such as in eq.~\eqref{eq:geos-vel}, coarse-graining $\bu$ is equivalent to coarse-graining each of its Cartesian components. 
We therefore transform the vector from spherical $(u_r,u_\lambda,u_\phi)$ to planetary Cartesian coordinates $(u_x,u_y,u_z)$ via:
\begin{linenomath*}
\begin{align}
u_x &= u_r \cos(\lambda)\cos(\phi) -u_\lambda \sin(\lambda) -u_\phi \cos(\lambda)\sin(\phi) \nonumber\\
u_y &= u_r \sin(\lambda)\cos(\phi) +u_\lambda \cos(\lambda) -u_\phi \sin(\lambda)\sin(\phi)\\
u_z &= u_r \sin(\phi) +u_\phi \cos(\phi)\nonumber
\lb{eq:Sph2Cart}\end{align}
\end{linenomath*}
where $\lambda$, $\phi$ are longitude and latitude, respectively, and $u_\lambda$, $u_\phi$ are the zonal and meridional velocity components, respectively. 
The radial velocity component, $u_r=0$ for the geostrophic flow. 
The conversion to Cartesian velocity components is necessary since the basis vectors for spherical velocities depend on space, while the Cartesian velocity basis vectors are spatially independent. 
Figure~\ref{fig:VelBasisVectors}, illustrates the spatial dependence of the velocity basis vectors. 
We apply the spherical convolution operation in eq.~\eqref{def:filtering} to each of $u_x$, $u_y$, $u_z$ as scalar fields to obtain the corresponding coarse-grained fields $\OL{u_x}$, $\OL{u_y}$, $\OL{u_z}$, then retrieve the coarse-grained velocity, $\OL{\bu}_\ell$ in spherical coordinates via 
\begin{linenomath*}
\begin{align}
\text{coarse radial flow}&= \OL{u_x} \cos(\lambda)\cos(\phi)  +\OL{u_y} \sin(\lambda)\cos(\phi)+\OL{u_z} \sin(\phi)=0\nonumber\\
\text{coarse zonal flow}&= -\OL{u_x} \sin(\lambda) +\OL{u_y} \cos(\lambda) \\
\text{coarse meridional flow}&= -\OL{u_x} \cos(\lambda)\sin(\phi)  -\OL{u_y} \sin(\lambda)\sin(\phi)+\OL{u_z} \cos(\phi).\nonumber
\lb{eq:Cart2Sph}\end{align}
\end{linenomath*}
That the `coarse-grained radial flow' (i.e. vertical flow, parallel to gravity) vanishes is not obvious and was proved in \citeA{aluie2019convolutions} and demonstrated numerically in \citeA{AluieTeeraratkul2023}. We emphasize that the coarse-graining algorithm we just described is valid only for non-divergent vectors such as $\bu$ in eq.~\eqref{eq:geos-vel}. Significant errors can arise for a general flow field \cite{AluieTeeraratkul2023}, where the complete coarse-graining formalism of \citeA{aluie2019convolutions} is necessary.

We use the coarse-graining kernel
\begin{linenomath*}
\begin{equation}
    G_{\ell}(\bx) = \frac{A}{2}\left( 1 - \tanh\left( 10\left (\frac{\gamma(\bx)}{\ell/2}-1 \right ) \right) \right),
    \label{eq:tanh-filter}
\end{equation}
\end{linenomath*}
%as shown in Figure \ref{fig:TanhKernel}. 
which is essentially a top-hat kernel \cite{pope2001turbulent} with graded edges. %to avoid numerical artifacts from the non-uniform discrete grid on the sphere. 
We use geodesic distance, $\gamma(\bx)$, between any location ${\bx=(\lambda,\phi)}$ on Earth's surface relative to location $(\lambda_0,\phi_0)$ where coarse-graining is being performed, which we calculate using
\begin{linenomath*}
\begin{equation}
    \gamma(\bx) = R_{\text{\tiny{Earth}}}\arccos\Big[\sin(\phi)\sin(\phi_0)+\cos(\phi)\cos(\phi_0)\cos(\lambda-\lambda_0)\Big].
    \label{eq:HaversineDistance}
\end{equation}
\end{linenomath*}
with ${R_{\text{\tiny{Earth}}}=6371~}$km for Earth's radius. 
In eq.~\eqref{eq:tanh-filter}, $A$ is a normalization factor, evaluated numerically, to ensure $G_\ell$ area integrates to unity as per equation \eqref{eq:normalization-condition}.
In general, we are not restricted to this choice of kernel; however, we use it because of its well-defined characteristic width $\ell$. Indeed, a convolution with $G_\ell$ in equation \eqref{eq:tanh-filter} is a spatial analogue to an  $\ell$-day running time-average (e.g., see Section \ref{sec:Spatio-temporalDecomposition}).

\subsubsection{Reflected hemispheres}
A basic complication that can arise when considering very large filter scales is that the filter may become incongruous with studying a smaller sub-domain. 
In this work, we are primarily concerned with the extra-tropical hemispheres: \([-90^\circ\mathrm{N},-15^\circ\mathrm{N}]\) and \([15^\circ\mathrm{N},90^\circ\mathrm{N}]\).
However, at very large length scales information from the equatorial band and opposing hemisphere can become introduced through an expanded filter kernel.
To resolve this issue, a `reflected hemispheres' approach is used, wherein one hemisphere is reflected and copied onto the other hemisphere, essentially producing a world with two north, or two south hemispheres.
This is the same methodology used in our previous work \cite{Storer2022}.

It is worth noting that the reflected hemispheres and equatorial masking would not be necessary in a context where non-geostrophic velocities are considered and a global power spectrum is desired.
They are used here because we wish to disentangle the power spectra of the geostrophic flow in the North and South.

\subsection{Partitioning the geostrophic kinetic energy}

From the coarse-grained horizontal geostrophic velocity field, $\OL{\bu}_\ell(\bx,t)$, following equation \eqref{def:filtering} as prescribed in \cite{aluie2019convolutions}, we partition kinetic energy (KE) into different sets of length-scales: 
\begin{linenomath*}
\begin{align}
\ME &= \frac{1}{2} |\bu(\bx,t)|^2 &\mbox{\text{(bare KE)}} \label{eq:tot_kene}\\
\ME_\ell &= \frac{1}{2} |\OL{\bu}_\ell(\bx,t)|^2 &\mbox{\text{(coarse KE)}}
\label{eq:coar_kene}\\
\ME_{< \ell} &= \frac{1}{2} \left (\OL{|\bu(\bx,t)|^2_\ell} - |\OL{\bu}_\ell(\bx,t)|^2 \right ) &\mbox{\text{(fine KE).}} \label{eq:fine_kene}
\end{align}
\end{linenomath*}
The ``bare KE'' in equation \eqref{eq:tot_kene} is the KE per unit mass (m$^2$/s$^2$) of the original geostrophic flow that includes all scales;  ``coarse KE'' in equation \eqref{eq:coar_kene} represents energy of the coarse-grained geostrophic flow at length-scales larger than $\ell$; and ``fine KE'' in equation \eqref{eq:fine_kene} accounts for geostrophic energy at scales smaller than $\ell$, which we discuss more in the following two paragraphs.
Partitioning geostrophic energy across scales is not trivial since one needs to ensure that such quantities are physically valid in the sense described by  \citeA{germano1992turbulence} and \citeA{vreman1994realizability}. In particular, it is important to ensure that the partitioned kinetic energy is (i) positive semi-definite ($\ge0$) at every $\bx$ and every time, and (ii) that summing the partitions yields the total energy.

While it is clear that ${\ME_\ell\ge0}$ in equation \eqref{eq:coar_kene}, this property is not obvious for $\ME_{< \ell}$ in equation \eqref{eq:fine_kene}. Moreover, it may not be obvious why $\ME_{< \ell}$ should represent energy at scales smaller than $\ell$.   \citeA{vreman1994realizability} showed that ${\ME_{< \ell}\ge0}$ if ${G_\ell\ge0}$, whereas $\ME_{< \ell}$ can be negative if the coarse-graining kernel $G_\ell$ is not positive semi-definite. 
A proof using convexity of the square function, $(\dots)^2$, illustrates why the first term $\OL{|\bu(\bx,t)|^2}_\ell$ in 
eq. \eqref{eq:fine_kene} has an overbar rather than defining fine KE as ${(|\bu(\bx,t)|^2 - |\OL{\bu}_\ell(\bx,t)|^2)/2}$. 
The proof from \citeA{sadek2018extracting} is as follows. When using ${G_\ell\ge0}$, coarse-graining $\OL{(\dots)}_\ell$ is a local averaging operation. From Jensen's inequality \cite{LiebLoss2001}, we know that ${\OL{[\mathcal{F}(\bu)]}_\ell\ge \mathcal{F}(\OL{\bu}_\ell)}$ for any convex operation, $\mathcal{F}$. Since ${\mathcal{F}(\bu) = |\bu|^2}$ is convex, we are guaranteed that ${\OL{|\bu(\bx,t)|^2_\ell} \ge |\OL{\bu}_\ell(\bx,t)|^2}$ and, therefore, ${\ME_{< \ell}\ge0}$ if the kernel ${G_\ell(r)\ge0}$, which is the case for our study (see equation \eqref{eq:tanh-filter}).

Regarding condition (ii) on the sum of energy partitions, \citeA{aluie2019convolutions} proved that (for a normalized $G_\ell$) the coarse-graining operation on the sphere in equation \eqref{def:filtering} preserves the spatial average of any field, ${\{\OL{F}_\ell(\bx)\} =\{F(\bx)\}}$, where ${\{\dots\} = (\text{Area})^{-1}\int \mathrm{d}\mathcal{S} (\dots)}$. Therefore, we have ${\left \{ \OL{|\bu|_\ell^2} \right \} = \left \{ |\bu|^2 \right \}}$.
This property guarantees that the sum of coarse KE and fine KE yields the total kinetic energy after integrating in space and in the absence of land,
\begin{linenomath*}
\be
\left \{ \ME \right \}= \left \{ \ME_\ell \right \} +\left \{ \ME_{< \ell} \right \}.
\label{eq:aver_cg-ener}
\ee
\end{linenomath*}
Eq. \eqref{eq:aver_cg-ener} justifies our interpretation of $\ME_{< \ell}$ as energy at scales smaller than $\ell$, since it is the difference between bare and coarse kinetic energy, on average, while also being positive locally. 

\subsection{Treatment of land-sea boundaries}
\label{sect:landtreatment}

In the above decomposition of energy, a choice has to be made in the presence of land. 
\citeA{Storer2022} provides some discussion on the subject, while here we discuss three possibilities, along with their pros and cons, in more detail.

\subsubsection*{Deformed kernel}

The `deformed kernel' approach is realized by coarse-graining ocean points near land with a kernel that is deformed or masked to avoid overlapping with land points. Such a deformed kernel must be renormalized to yield an average over just ocean points rather than the whole sphere. The main advantage of this approach is that it treats land as a well-defined boundary that is separate from the ocean regardless of the coarse-graining length-scale. It is also familiar to ocean modelers who routinely mask values over land and do not include such masked values when performing area averages. 

However, the deformed kernel has disadvantages that motivate against its use for coarse-graining ocean flows. 
First, a kernel that is inhomogeneous (\textit{i.e.} changes shape depending on geographic location) does not conserve domain averages, including the kinetic energy of the flow. 
The reason for this failed conservation is detailed in  \ref{appendix:deforming_kernel} and demonstrated in Figure \ref{fig:energy-leak} (blue plot). 
This figure shows how a kernel that is deformed (via masking) to exclude land does not yield $100\%$ of the total energy, \textit{i.e.,} it does not satisfy equation \eqref{eq:aver_cg-ener}. As a result, it can yield total energy that is either less than $100\%$  (e.g., over scales larger than $500~$km in Figure \ref{fig:energy-leak}) or  \emph{greater} than $100\%$ (e.g., between $100~$km and $400~$km in Figure \ref{fig:energy-leak}). 

For some purposes, the total energy values in Figure \ref{fig:energy-leak} are fairly close to $100\%$ (deviations less than $1\%$) so one  might argue that the deformed kernel is suitable in practice. Nonetheless, a more basic reason to avoid deformed kernels is that such inhomogeneous kernels (which also include averaging values at adjacent grid-cells or block-averaging on the sphere) do not commute with spatial derivatives. Consequently, the coarse-grained field resulting from a deformed kernel is not guaranteed to satisfy fundamental flow properties exhibited by the unaveraged flow, such as non-divergence, geostrophy, and the vorticity present at various scales. These considerations are further detailed in \citeA{aluie2018mapping} and \citeA{aluie2019convolutions}.

\subsubsection*{Fixed kernel}

The `fixed kernel', also used in Figure \ref{fig:energy-leak}, is homogeneous so that it preserves its shape at all locations. When coarse-graining ocean points near land such that the kernel overlaps land points, we treat land points in a manner consistent with the boundary conditions between land and ocean. For example, if we are coarse-graining the velocity, we treat land as water with zero velocity, which is consistent with the formulation of OGCM where land is often treated as a region of zero velocity. Furthermore, we include these zero land values as part of the coarse-graining operation. 

This choice may seem unnatural since we are including unphysical values within the coarse-graining operation. However, it is helpful to think of coarse-graining as an operation analogous to removing one's eyeglasses, rendering an image fuzzy and boundaries less well-defined. When coarse-graining at a scale $\ell$, the precise boundary between land and ocean becomes blurred at that scale and its precise location becomes less certain. The coarse-grained
velocity, $\OL\bu_\ell$, can be nonzero within a distance $\ell/2$ beyond the continental boundary over land. Forfeiting exact spatial localization in order to gain scale information is theoretically inevitable due to the uncertainty principle, which prevents the simultaneous localization of data in physical-space and in scale-space \cite{SteinWeiss,Sogge}. The main advantage of the ``Fixed Kernel'' choice is ensuring that coarse-graining and spatial derivatives commute so that it  preserves the fundamental physical properties (symmetries) of the flow. Further discussion of these issues can be found in \citeA{aluie2018mapping} and \citeA{aluie2019convolutions}.

\subsubsection*{Fixed kernel with or without land}

\begin{figure}[ht]
    \centering
    \includegraphics[scale=1]{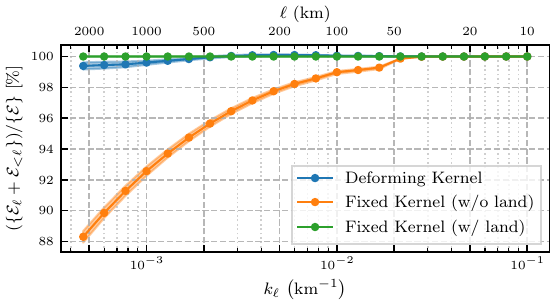}
    \caption{
      Percentage of total energy recovered by summing the fine and coarse KE terms in equation \eqref{eq:aver_cg-ener} obtained by coarse-graining over the full ocean surface as a function of the filter scale, ${k_{\ell}=1/\ell}$. 
        The three lines correspond to the three approaches described in section \ref{sect:landtreatment}, namely, filtering with a fixed kernel shape and excluding/including land (orange/green lines) when tallying the total energy. We also coarse-grain with a deformable filter kernel to exclude the filter overlapping land regions (blue line).
        }
    \label{fig:energy-leak}
\end{figure}

After coarse-graining the velocity field with a fixed kernel, we show in Figure \ref{fig:energy-leak} the level of energy conservation if we include or exclude land points from the final tally of kinetic energy. We call these, respectively, the `fixed kernel w/ land' and `fixed kernel w/o land'. 
The latter (orange line) highlights how coarse-graining smears energy onto land (within $\ell/2$ distance inland) such that if we exclude land from the final tally, we find some leakage of energy onto land, which increases as the coarse-graining scale $\ell$ increases. 
We find energy leakage of the order of $1\%$ at coarse-graining scales $<100~$km, $\approx4\%$ for scales $\lesssim500~$km, and up to $12\%$ at scales of order $2000~$km. 
However, if we choose to include land in our final tally, we are guaranteed to conserve $100\%$ of the energy by satisfying equation \eqref{eq:aver_cg-ener}, thus ensuring that the energy budget is fully closed. 
After all, in an ocean model on a discrete grid, the land boundary is only expected to be accurate within a $\Delta x$ distance from any estimate of the truth, where $\Delta x$ is analogous to our coarse-graining scale $\ell$. 

\subsubsection*{What we use here}

While we have implemented all three approaches to coarse-graining, unless otherwise stated in this work, we choose the fixed kernel w/ land by including land regions that have non-zero velocity (again, as realized through leakage from nearby ocean values). \citeA{Storer2022} showed that deformed and fixed kernels yield qualitatively consistent results for spectra.
We avoid coarse-graining with a deformed kernel to remain consistent with previous work \cite{aluie2018mapping} and with forthcoming studies where we apply coarse-graining to the dynamical equations where commuting with spatial derivatives is essential.

\subsection{The filtering spectrum}
\label{sect:filt_spectr}

\citeA{sadek2018extracting} showed how coarse-graining can be used to extract the energy content at different length scales. They do so by partitioning the velocity into discrete length scale bands rather than the two sets (coarse KE and fine KE) in equations \eqref{eq:coar_kene} and \eqref{eq:fine_kene}. The resulting quantity is called the \textit{filtering spectrum}.
The filtering spectrum is distinct from the traditional Fourier spectrum, with  coarse-graining offering a way to measure energy distributions without relying on a Fourier transform, thus avoiding the limitations noted in Section \ref{sec:Fourier}.  

The filtering spectrum is obtained by differentiating in scale the coarse KE
\begin{linenomath*}
\be
\olE(k_\ell) = \frac{d}{dk_\ell} \left \{ \ME_{\ell} \right \} = -\ell^2 \frac{d}{d \ell} \left \{ \ME_\ell \right \},
\label{eq:filt_spect}
\ee
\end{linenomath*}
where ${k_\ell = 1/\ell}$ is the `filtering wavenumber'. \citeA{sadek2018extracting}  showed that the filtering spectrum satisfies energy conservation and that ${\olE(k_\ell,t) \ge 0}$ when using certain types of kernels (e.g., concave) of which the top-hat kernel is an example. Moreover, \citeA{sadek2018extracting} identified the conditions on $G_\ell$ for $\olE(k_\ell,t)$ to be meaningful in the sense that its scaling agrees with that of the traditional Fourier spectrum (when a Fourier analysis is possible, such as in periodic domains).
Below, we shall sometimes refer to $\ME_\ell$ as the `cumulative spectrum' following \citeA{sadek2018extracting} since it accounts for all energy at scales larger than $\ell$. In contrast, $\olE(k_\ell,t)$, is the spectral energy \emph{density} at a specific scale $\ell$.

\subsection{ Comparison with Spherical Harmonics }
\label{sec:spherical_harmonics}
Our previous results on spectra using CG in \citeA{Storer2022} provide justification for using spherical harmonics on the global ocean and a guide for treating land in a manner that is consistent with boundary conditions. For the ocean velocity, the boundary conditions are zero normal velocity (no flow through) and zero tangential velocity (no-slip). Therefore, when using spherical harmonics, we set land to have zero velocity values, similar to what we do with the CG method.

\begin{figure}
    \centering
    \includegraphics[scale=1]{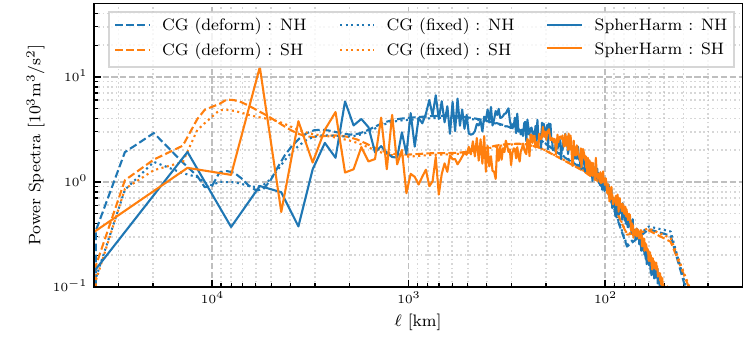}
    \caption{ 
        \textbf{ Power Spectra with Spherical Harmonics and Coarse-Graining }
        Power spectra computed using spherical harmonics (solid lines), coarse-graining with a deforming kernel (dashed lines), and coarse-graining with a fixed kernel (dotted lines).
        Reflected hemispheres were used to obtain spectra for NH and SH separately.
        Note that these spectra were obtained by masking out only a thin strip \([2^\circ\mathrm{S},2^\circ\mathrm{N}]\) and integrating over the domain to allow for the application of spherical harmonic transforms, unlike those of Figure~\ref{fig:spectra} and \cite{Storer2022} that only integrated over latitudes outside of \([15^\circ\mathrm{S},15^\circ\mathrm{N}]\), explaining the discrepancy in peak locations. 
    }
    \label{fig:spher_harmonic:spectra}
\end{figure}

Figure~\ref{fig:spher_harmonic:spectra} compares spectra from CG to those from spherical harmonics.  It uses a single daily mean of AVISO data using spherical harmonics, coarse-graining with a deforming kernel, and coarse-graining with a fixed kernel.
The spherical harmonic analysis was performed using PySHTools \cite{Wieczorek2018} on the AVISO data with reflected hemispheres.

The two methods yield qualitatively consistent domain-averaged results, such as the broad mesoscale peak, the NH gyre peak, and the ACC peak. Both spectra (spherical harmonics and CG) integrate to the same total energy. However, the spherical harmonics spectra are too noisy at gyre-scales ($>1000~$km). At these large length-scales (low modes), spherical harmonics spectra have poor scale resolution because the eigenmodes are spaced far apart; in integer multiples of the fundamental mode. It is particularly noticeable around the ACC peak at $\ell\approx10^4$ km. This limitation is shared by Fourier methods in a Cartesian box. This is not a limitation for the CG method of computing spectra since it conserves energy without relying on the orthogonality structure of an eigenbasis in the strict sense \cite{sadek2018extracting}.

\begin{figure}
    \centering
    \includegraphics[scale=1.]{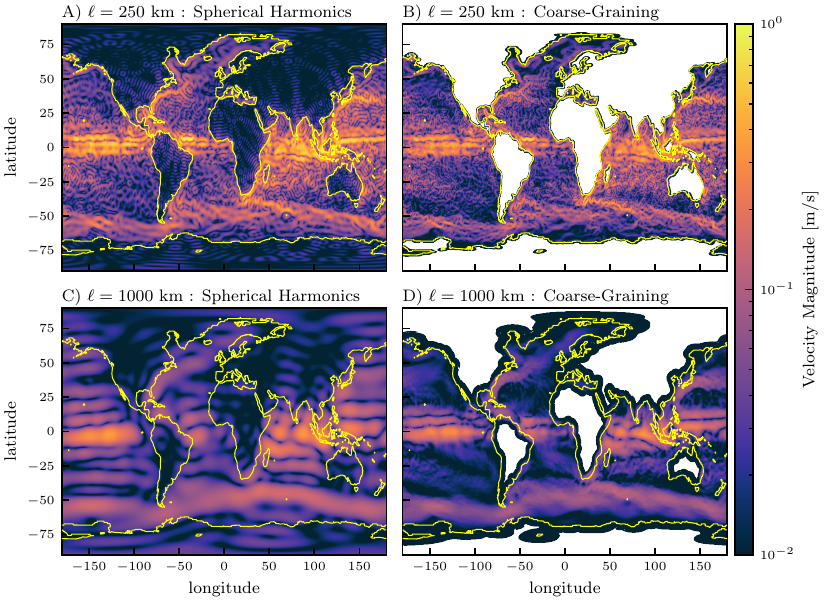}
    \caption{ 
        \textbf{ Filtering Maps with Spherical Harmonics and Coarse-Graining }
        Speed of the large-scale AVISO surface currents obtained by \textbf{[left, AC]} spherical harmonics and \textbf{[right, BD]} coarse-graining.
        Velocity fields are filtered at \textbf{[top, AB]} 250~km and \textbf{[bottom, CD]} 1000~km.
        Colour maps show velocity magnitude on a logarithmic scale, with white indicating identically zero values.
        Yellow contours indicate land boundaries in the unfiltered data. Note how filtering with spherical harmonics, even at $250~$km, yields non-zero flow over all continents and prominent ringing patterns. This is due to the inherently global nature of spherical harmonics, which makes it challenging to infer spatially local information at different scales.
    }
    \label{fig:spher_harmonic:maps}
\end{figure}

A main disadvantage of spherical harmonics is that they are inherently global and cannot provide local information connecting scales with currents geographically.
This becomes apparent in spatial maps, such as those in Figure~\ref{fig:spher_harmonic:maps}.
In coarse-graining, non-zero current velocities only intrude a distance of $\ell/2$ inland from the coast, as evidenced by the thin band of dark colours inside the yellow contour lines (coastlines).
Moreover, the band within the yellow contour is dark, which reinforces that very little energy is distributed over land. Even at a 1000~km filter scale, the majority of land retains identically zero velocity, indicated by white.
In contrast, even at a small filter scale, spherical harmonics generate beams of spectral ringing that extend deep into land regions, with non-trivial magnitudes.
Worse still, at a 1000~km filter, the spherical harmonic filtering fills the global ocean with zonal bands, even in the more quiescent open oceans.
These ringing features are not present under a coarse-graining approach with an appropriately chosen kernel.

In addition, there are practical considerations in regards to comparing coarse-graining with spherical harmonics.
Like traditional Fourier methods, spherical harmonics require the input data to conform to fairly strict structures: uniform lat/lon grids, specific resolution aspect ratios, etc.
In contrast, coarse-graining is grid agnostic.
That is, while the implementation details are different, coarse-graining applies just as well to a uniform lat/lon grid as to a generalized non-uniform triangularization grid.
While FlowSieve (\url{https://github.com/husseinaluie/FlowSieve}), the coarse-graining package used in this work, at present only accepts rectangular (but non-uniform) lat/lon grids, that is a limitation imposed by the current implementation, and not by the underlying methodology.

\subsection{Reynolds averaging}
\label{sect:meth-RA}

We close this section by reviewing basic properties of Reynolds averaging (RA) as realized by time averages.

\subsubsection*{Basics of Reynolds averaging}

Time averaging separates the flow into a  time-average/`mean' and a fluctuating/`eddy' as given by \cite{pope2001turbulent}
\begin{linenomath*}
\be
\langle \bu \rangle (\bx) = \frac{1}{T} \int_{t_0}^{t_0+T} \bu(\bx,t) \mathrm{d}t,
\label{eq:mean_vel}
\ee
\end{linenomath*}
\begin{linenomath*}
\be
\bu'(\bx,t) = \bu(\bx,t) - \langle \bu \rangle (\bx),
\label{eq:eddy_vel}
\ee
\end{linenomath*}
where $\langle \bu \rangle$ is the mean component, $\bu'$ the eddy component, and $T$ represents the entire time record and not just a time window. Two key properties of the Reynolds decomposition are 
\begin{equation}
\langle\langle \bu \rangle\rangle = \langle \bu \rangle
\quad \mbox{and} \quad 
\langle \bu' \rangle = 0,
\end{equation}
so that the mean of a mean returns the mean (idempotent property) while the mean of the eddy is zero. The resulting mean and eddy kinetic energy components are respectively given by 
\begin{linenomath*}
\be
MKE (\bx) = \frac{1}{2} |\langle \bu\rangle|^2(\bx),
\label{eq:mean_ene}
\ee
\be
EKE(\bx,t) = \frac{1}{2} |\bu'|^2(\bx,t).
\label{eq:eddy_ene}
\ee
\end{linenomath*}
Notice that the sum of mean and eddy kinetic energy is not equal to the total kinetic energy. Rather, there is an extra cross term, ${\bu' \cdot \langle \bu\rangle}$, needed to close the budget. However, the cross term is not positive definite and it has a zero time average, ${\langle \bu' \cdot \bu \rangle =0}$. 
Following a RA decomposition, the total energy can be written as
\begin{linenomath*}
\be
\ME(\bx,t) = EKE(\bx,t) + MKE(\bx) + \frac{1}{2}\left (\bu' \cdot \langle \bu\rangle \right )(\bx,t).
\label{eq:tot_ene2}
\ee
\end{linenomath*}

\subsubsection*{Key differences between Reynolds averaging and coarse-graining}

A key difference between coarse-graining and Reynolds-averaging is that within RA, applying the averaging operation twice on any field yields the same result whereas that property does not hold for coarse-graining with non-projector kernels \cite{buzzicotti2018effect}:
\begin{linenomath*}
\be
\langle\langle F \rangle\rangle =\langle F \rangle
\quad \mbox{whereas} \quad 
\OL{\OL{F}} \ne \OL{F},
\ee
\end{linenomath*}
where \({<\cdot>}\) denotes time averaging and \(\overline{\,\cdot\,}\) denotes coarse-graining.
Another important difference is that a Reynolds average does not provide a control to adjust the partition between the `mean' and `eddy' components. That is, a Reynolds decomposition is not a length-scale decomposition and this point is illustrated in section \ref{sec:Spatio-temporalDecomposition}.
Consequently, the time-mean or ensemble-mean flow is not synonymous with large-scale flow, nor does a Reynolds eddy fluctuation directly correspond to a characteristic fine-scale. 

To help understand the above points, we emphasize the distinction between  time-scale and decorrelation-time for a particular flow feature. While it is generally true that larger (smaller) scales have slower (faster) time-scale dynamics, it is not always true that their decorrelation-time follows this relation. As an example, consider  stationary eddies, such as the Mann eddy in the North Atlantic. Such eddies have a small spatial-scale (relative to the gyre or basin) but are persistent in time. 
As a result, even if the timescale (${\sim \ell/u}$) for a structure is small when it is associated with the relatively fast dynamics of eddying flows, it can be highly correlated (or even stationary) in time, so that its contribution to the $MKE$ is not completely removed by a time-average. We show this behavior in sections \ref{sec:Spatio-temporalDecomposition} and \ref{sec:Spatio-temporal}.

\section{Analysis results}
\label{sect:Results}

In this section we present results of the coarse-graining analysis along with a comparison with Reynolds averaging based on time averages. In the second part of this section we present results from coarse-graining in both space and time as a means to characterize the time-scales associated with different length-scales.

\subsection{Coarse-graining the surface geostrophic flow from AVISO}

\begin{figure}[ht]
    \centering
    \includegraphics[scale = 1]{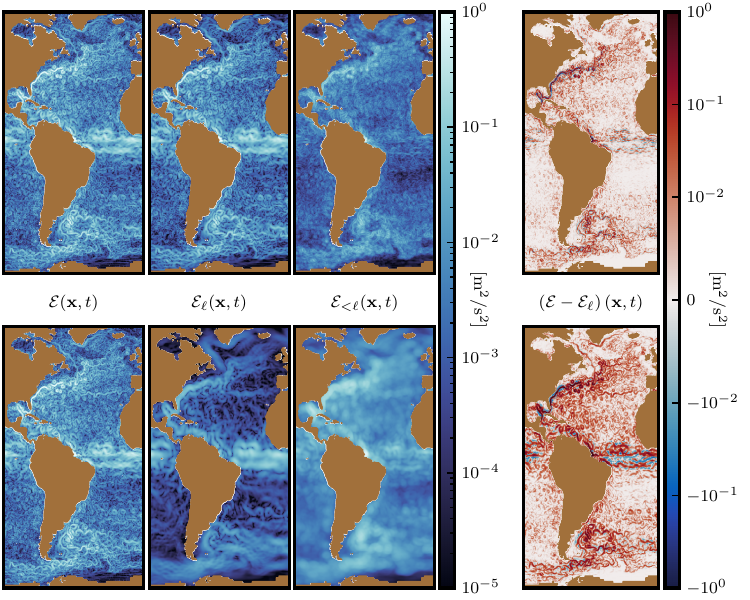}
    \caption{
        Maps of the coarse-grained decomposition of kinetic energy from a single day of the AVISO analysis at two different filter scales, ${\ell=100}$ km (top) and ${\ell=400}$ km (bottom). Here the bare KE, $\ME(\bx,t)$, is compared with coarse KE, $\ME_\ell (\bx,t)$, and fine KE, $\ME_{<\ell}(\bx,t)$. 
        The right-most column shows the fine scale term defined by equation \eqref{eq:2fine_kene}, which can yield negative values.
    }
    \label{fig:coar-gra-maps}
\end{figure}

We split the geostrophic kinetic energy from AVISO into its fine and coarse-grained components following equations \eqref{eq:coar_kene} and \eqref{eq:fine_kene}. 
For a qualitative appreciation of this decomposition, Figure \ref{fig:coar-gra-maps} displays maps of the kinetic energy just over the Atlantic using two different filter scales, $\ell = 100~\mbox{km}$ in the top row and $\ell = 400~\mbox{km}$ in the bottom row. From left to right, panels in Figure \ref{fig:coar-gra-maps} show the total kinetic energy, $\ME$, the coarse energy, $\mathcal{E}_\ell$, and the fine energy, $\ME_{<\ell}$. 
The fine scale kinetic energy, $\ME_{<\ell}$, represents kinetic energy at scales less than $\ell$, as  represented (or projected) on a grid of resolution ${\Delta x \sim \ell}$. Notably, as seen in Figure \ref{fig:coar-gra-maps}, $\ME_{<\ell}$ does not have small scale features, which results since there is a filter applied to both terms in equation \eqref{eq:fine_kene} defining $\ME_{<\ell}$. This definition ensures that $\ME_{<\ell}$ is positive semi-definite at each point in space and time. 

Visualization of fine kinetic energy, $\ME_{<\ell}$,  is still useful to identify the regions where structures smaller than the filter scale are dominant in the ocean. Even so, one may wish to view the alternative quantity 
\begin{linenomath*}
\be
\ME - \ME_{\ell} = \frac{1}{2} \paren{ \absv{\bu(\bx,t)}^2 - \absv{\OL{\bu}_\ell(\bx,t)}^2 },
\label{eq:2fine_kene}
\ee
\end{linenomath*}
which is shown in the right-most column of Figure~\ref{fig:coar-gra-maps}. 
This quantity reveals more fine scale features since only the second term on the right hand side is filtered.  However, as discussed in Section \ref{sect:meth-CG}, the energy difference, ${\ME - \ME_{\ell}}$, can be negative locally in space, and so it does not serve our purposes for decomposing the energy into non-negative terms.

\subsection{Reynolds averaging decomposition}

\begin{figure}[ht]
    \centering
    \includegraphics[scale = 1]{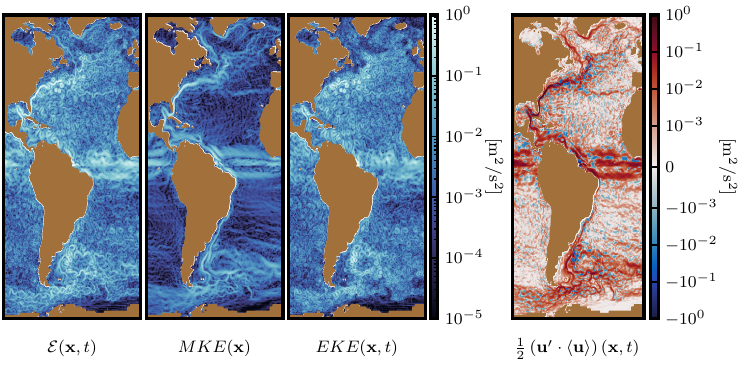}
    \caption{
    Decomposition of geostrophic kinetic energy from AVISO for the Atlantic basin from a time averaging (Reynolds) decomposition. Left panel: total energy, $\ME(\bx,t)$ at a single day. Left middle  panel: 9-year time mean, $MKE(\bx)$.  Right middle panel: fluctuating eddy term, $EKE(\bx,t)$. Right panel: the cross term required to recover the total geostrophic energy as defined in equation~(\ref{eq:tot_ene2}). 
    }
    \label{fig:RA-maps-AVISO}
\end{figure}

Here, and in subsequent subsections, we show that the time-mean flow consists of an entire range of length scales with substantial contributions from the mesoscale. Figure~\ref{fig:RA-maps-AVISO} shows the mean-fluctuation decomposition following the Reynolds averaging approach. The maps are focused on the Atlantic region to help reveal details and we show just those obtained from AVISO.
The time mean is obtained by averaging the velocity over the whole time series available, spanning nine years. From left to right we show the total energy at a single day, the time mean energy, $MKE(\bx)$, the fluctuating eddy term, $EKE(\bx,t)$, and the cross term, $(\bu' \cdot \langle \bu\rangle )/2$.

Having used a relatively long time series for averaging, the mean energy in Figure \ref{fig:RA-maps-AVISO} is rather depleted away from major current systems, so that the Gulf Stream and the Antarctic Circumpolar Current are quite pronounced relative to the gyre interiors. We  appreciate from this figure that the mean flow retains a substantial contribution from structures with a variety of sizes. 
In the same way, the `eddy' (or temporally fluctuating) flow in Figure~\ref{fig:RA-maps-AVISO} contains most of the small scale fluctuations but also a substantial contribution from large-scale structures. 
The cross term shown on the right panel of Figure~\ref{fig:RA-maps-AVISO}  has strong fluctuations around zero, which make its contribution almost (but not exactly) zero after a spatial-average. 
The blending of length scales revealed by these figures reflects the inability of time averaging to decompose the kinetic energy according to length-scales. 

\begin{figure}[ht]
    \centering
    \includegraphics[scale =1]{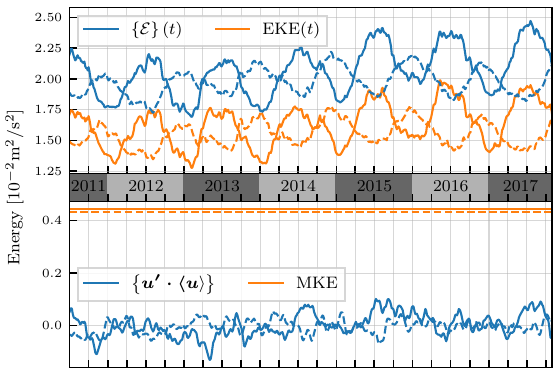}
    \caption{
        Top panel: Time-series of total geostrophic kinetic energy, $\left \{\ME(\bx) \right \}(t)$ (blue), and the fluctuating component, $\left \{EKE(\bx) \right \}(t)$ (orange), in the North (solid line) and South (dashed line) from the AVISO analysis. Vertical grid lines indicate the start of each quarter-year (01Jan, 01Apr, 01Jul, 01Oct).
        Bottom panel: Time-series of the cross term (blue) and kinetic energy of the 9-year mean, $\{MKE(\bx)\}$ (orange), in the North (solid line) and South (dashed line). $EKE$ constitutes a substantial portion of the total energy and with an almost indistinguishable temporal variation. Here, we show only 6.5 years of the full 9-year record. Plots shown use a 4-day sampling frequency, but averages are based on a 1-day sampling of the 9-year record.
        }
    \label{fig:RA-AVISO-Season}
\end{figure}

To further investigate the role of the three Reynolds average energy terms,  Figure~\ref{fig:RA-AVISO-Season} shows their temporal variability in both hemispheres. 
In the first row, we see that $EKE$ constitutes a substantial portion of the total energy $\ME$ ($80\%$) and their temporal evolution is almost indistinguishable. Both $EKE$ and $\ME$ tend to peak during the spring-summer.
The bottom row of Figure~\ref{fig:RA-AVISO-Season} shows $MKE$, which is independent of time, and the cross term, which has a zero average. These two quantities are much less energetic, with the mean term ${\approx20\%}$ of the total and the cross term fluctuates about its zero average without a clear seasonal signal.

\subsection{The filtering spectrum}
\label{sect:filt_spectr2}

\begin{figure}
    \centering
    \includegraphics[scale=1]{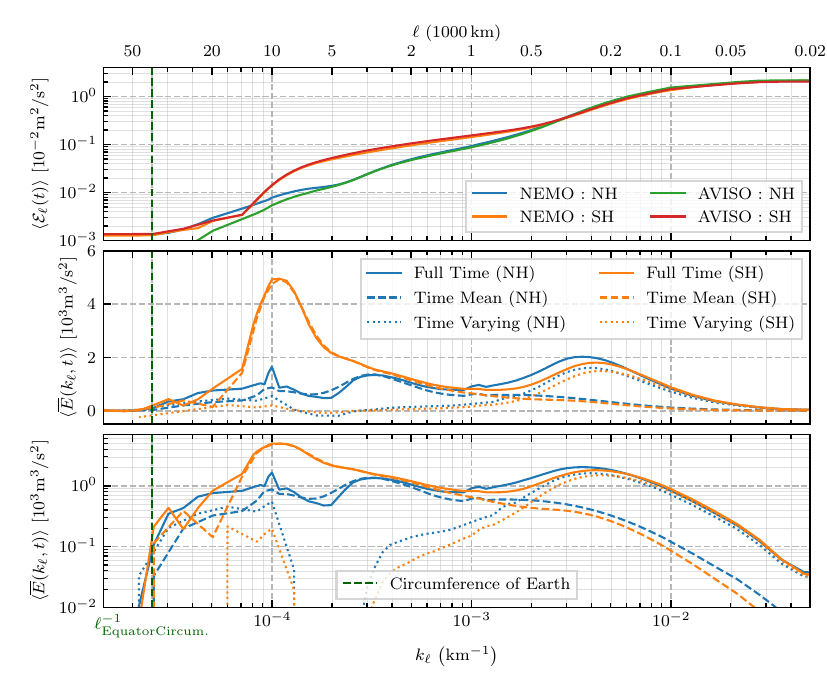}
    \caption{ 
        \textbf{Power Spectra}
        \textbf{[Top]}: Cumulative surface geostrophic kinetic energy spectra, $\ME_\ell$, as a function of scale $\ell$, obtained from both the AVISO and NEMO products in the North and South. 
        \textbf{[Middle and bottom]}: Filtering spectra obtained following eq.~\eqref{eq:filt_spect} for the full (solid lines), time mean (dashed times), and time-varying (dotted liens) ssh-derived geostrophic velocity from the NEMO dataset.
        Note that both middle and bottom panels show the same data, but using lin-log and log-log scales respectively.
    }
    \label{fig:spectra}
\end{figure}

In Figure~\ref{fig:spectra} we show the cumulative large-scale energy for the north and south hemispheres as obtained from equation~\eqref{eq:filt_spect} for AVISO and NEMO, as well as the filtering spectra for the Reynolds-decomposed components of NEMO: full time signal, $\ME(\bx,t)$, time mean, $MKE(\bx)$, and time varying, $EKE(\bx,t)$.
In the top panel we show the cumulative area-averaged energy spectra, $\ME_\ell$, as a function of coarse-graining scale. 
In the centre and bottom panels, we show the filtering spectrum (c.f. equation \eqref{eq:filt_spect}), in lin-log and log-log scale respectively. 

\paragraph*{Cumulative Energy Spectra}
At the large $k_\ell$ (small \(\ell\)) end of the cumulative spectra, we see that all four datasets converge.
That is, for both NEMO and AVISO, the area-averaged energy density is \({\approx2\times10^{-2}\mathrm{m}^2/\mathrm{s}^2}\) (corresponding to an RMS velocity of \({\approx20~}\)cm/s), for either hemisphere.
At gyre-scales, SH has noticeably higher energy density than NH. This asymmetry is balanced by an opposing asymmetry over the mesoscales, where NH has higher KE density, which is more readily detectable in the filtering spectra. The NH-SH asymmetry can be attributed to basin geometry and continental boundaries. The NH ocean basins are land-constrained relative to the SH, which has more room for a larger-scale flow, namely the ACC, to develop and intensify. We shall see in Table~\ref{table:ell_band_stats} below that most of the hemispheric asymmetry resides in the stationary time-mean flow. The stronger (energy-per-area) mesoscale flow in the NH is stationary and is most probably due to the time-invariant forcing exerted by continental boundaries. 
This can explain our observation in Fig.~\ref{fig:spectra} (middle panel) that NH mesoscales are more intense than in the SH.

\paragraph*{Filtering Spectra}
The full time filtering spectra in Fig.~\ref{fig:spectra} have been previously reported in \citeA{Storer2022}.
Here, we extend previous results by incorporating CG spectra of the time-mean and time-varying Reynolds averaging components.
As might be expected, the time-mean velocity peaks spectrally at large scales (\({\ell>10^3~}\)km), while the time-varying component peaks over the mesoscales. This may misleadingly suggest that time-averaging produces a scale separation to a good approximation.
However, as will be shown later in this subsection, the mesoscale energy (area under the spectrum plot) accounts for a majority of the time-mean energy.  Therefore, as we are going to show, the time-mean flow is dominated by stationary mesoscale structures $<500~$km in size.
The length-scale at which spectra of the time-varying and time-mean velocity cross is slightly larger than 500~km.

\paragraph*{ Proportion of Energy in Mesoscales }
In Table~\ref{table:ell_band_stats} we present the kinetic energy of the Reynolds averaging components partitioned at 500~km for the NEMO dataset.
There are three primary conclusions that can be drawn from Table~\ref{table:ell_band_stats}.
1)~While mesoscales are dominated by time-varying flow, the majority of the time-mean energy is also in the mesoscales.
2)~The geostrophic time-varying flow is nearly entirely mesoscale, with only a few percentage points in larger scales.
It is important to recall, however, that this analysis excludes ageostrophic motions, such as the Ekman flow.
3)~While the full and time-varying velocities are generally consistent between hemispheres, the time-mean velocity shows strong asymmetry. 
Specifically, time-mean mesoscles are stronger in NH, while time-mean gyre-scales are stronger in SH.
A likely contributor to the latter is the ACC, which contains large-scale time-mean currents.
In the NH, there is stronger stationary forcing at the mesoscales relative to the SH due to more restrictive continental boundaries. 
Nearly identical results are found from the Reynolds averaging decomposition applied over the 9-year AVISO dataset, shown in~\ref{appendix:ra_aviso}.

\begin{table}[ht]
    \centering
    \begin{tabularx}{0.9\textwidth}{|c|c|c|c|c|}
    \cline{1-5}
    \multicolumn{2}{|c|}{}  & Full Velocity & Time-Mean & Time-Varying
    \\ \cline{1-5}
    \multirow{2}*{ \(\ell<\) 500~km [\(10^{-2}\)m\(^2\)/s\(^2\)] } & NH & \phz2.1\phz & \phz0.36 & \phz1.7\phz 
    \\ \cline{2-5}
    & SH & \phz2.0\phz & \phz0.25 & \phz1.8\phz
    \\ \cline{1-5}
    \multirow{2}*{ \(\ell>\) 500~km [\(10^{-2}\)m\(^2\)/s\(^2\)] } & NH & \phz0.20 & \phz0.15 & \phz0.06 
    \\ \cline{2-5}
    & SH & \phz0.22 & \phz0.19 & \phz0.04
    \\ \cline{1-5}
    \multirow{2}*{ \(\ell<\) 500~km [\% of Total] } & NH & 91\phpz & 71\phpz & 97\phpz
    \\ \cline{2-5}
    & SH & 90\phpz & 57\phpz & 98\phpz
    \\ \cline{1-5}
    \end{tabularx}
    \caption{
        \textbf{ Mesoscale Energy for Reynolds' Components }
        The area-mean kinetic energy partitioned at 500~km for each hemisphere (equivalent to the top panel of Fig.~\ref{fig:spectra}), for the three Reynolds' components: full $\ME(\bx,t)$, time-mean $MKE(\bx)$, and time-varying velocity $EKE(\bx,t)$.
        Presented values are the median (50\(^{\mathrm{th}}\) percentile) in time from the NEMO dataset.
    }
    \label{table:ell_band_stats}
\end{table}

\paragraph*{RMS Velocity in Major Currents}

By integrating the filtering spectrum over a scale band, we can obtain the total KE for the chosen scale band and, subsequently, the RMS velocity for that range of spatial scales.
Table~\ref{table:rms_vels} presents these RMS velocity magnitudes from NEMO for a selection of geographic regions: NH, SH, ACC, Gulf Stream, and Kuroshio, both within the mesoscale (100--500~km) and gyre-scale (\({>10^3~}\)km) scale-bands.
The region definitions are included in \ref{appendix:region_defs}.
Note that mesoscales are stronger in NH than SH, while gyre-scales are stronger in SH.

\begin{table}[ht]
    \centering
    \begin{tabularx}{0.90\textwidth}{|c|c|c|c|c|}
    \cline{1-5}
    \multirow{2}*{Region} & \multicolumn{2}{c|}{Mesoscales (100--500~km)} & \multicolumn{2}{c|}{Gyre-scales (\(>10^3~\)km)}
    \\ \cline{2-5}
    & Block Region & KE Masked
    & Block Region & KE Masked
    \\ \cline{1-5}
    South of Tropics & 15.0 & --- & 5.3 & ---
    \\ \cline{1-5}
    ACC & 16.4 & 28.1 & 7.0 & \phz9.7
    \\ \cline{1-5}
    North of Tropics & 15.5 & --- & 4.3 & ---
    \\ \cline{1-5}
    Gulf Stream & 32.7 & 42.2 & 7.8 & \phz8.7
    \\ \cline{1-5}
    Kuroshio & 26.5 & 40.0 & 8.1 & 10.1
    \\ \cline{1-5}
    \end{tabularx}
    \caption{
        \textbf{ RMS Current Speed [cm/s] in Select Regions }
        The area-mean RMS velocity magnitude [cm/s] for selected regions using both Block and KE-masked definitions, see~\ref{appendix:region_defs}.
        Note that there is no KE-masked variant of the NH and SH regions.
        Reported values are for the time median (\(50^{\mathrm{th}}\) percentile).
        Presented values are from the NEMO dataset, and are all rounded to one decimal point.
    }
    \label{table:rms_vels}
\end{table}

\paragraph*{Extrapolating to Smaller Scales}
Both NEMO and AVISO datasets agree well on the spectral energy density of the mesoscales, down to \({\approx100~}\)km, where resolution effects begin to cause deviations \cite{amores2018up,ballarotta2019resolutions}.
Using \(k_{\ell}^{-3}\) and \(k_{\ell}^{-5/3}\) power laws, we can extend the power spectrum towards smaller scales.
Note that this is presented as a thought experiment, and is not intended to suggest that such a power law will hold over all smaller scales. 
If we let \(S_{100\mathrm{km}}\) denote the spectral energy density for \({\ell=100~}\)km, and assume a spectral scaling of \(k^{-\alpha}\) spanning all scales smaller than 100~km, then we can compute the total amount of energy in scales smaller than \(100~\)km as
\begin{linenomath*}
\be
\lim_{n\to\infty}\int_{k_\ell=10^{-5}}^{10^n} S_{100\mathrm{km}} 10^{-5\alpha}k^{-\alpha} \mathrm{d}k
=
\frac{1}{\alpha-1}S_{100\mathrm{km}}10^{-5},
\label{eq:extrapolate_KE}
\ee
\end{linenomath*}
or, alternatively, to only consider the decade spanning 10--100~km,
\begin{linenomath*}
\be
\int_{k_\ell=10^{-5}}^{10^{-4}} S_{100\mathrm{km}} 10^{-5\alpha}k^{-\alpha} \mathrm{d}k
=
\frac{1}{\alpha-1}S_{100\mathrm{km}}10^{-5}\left[1-10^{1-\alpha}\right],
\label{eq:extrapolate_KEv2}
\ee
\end{linenomath*}
where we assume that \({\alpha>1}\).
Using equations \eqref{eq:extrapolate_KE} and \eqref{eq:extrapolate_KEv2} and the 100~km values presented in Fig.~\ref{fig:spectra}, we can then compute the amount of energy in scales smaller than 100~km as a percentage of energy \emph{across all scales}.
These values are presented in Table \ref{table:sub100km_energy} and reveal that as much as 25--50\% of the surface geostrophic kinetic energy is contained in scales smaller than 100~km. These scales are un(der)-resolved by pre-SWOT satellite products.
Our estimates are contingent on a persistent power-law scaling over small scales, but they nevertheless illustrate how a substantial proportion of surface geostrophic energy may be missed by coarse resolution.
\begin{table*}[ht]
    \centering
    \begin{tabularx}{0.70\textwidth}{|c|c|c|c|c|}
    \cline{1-5}
    \multirow{2}*{\(-\alpha\)}& \multicolumn{2}{c|}{AVISO} & \multicolumn{2}{c|}{NEMO}
    \\\cline{2-5}
    & NH & SH & NH & SH
    \\\cline{1-5}\cline{1-5}
    \(-3\)            & 24\% [24\%] & 25\% [25\%] & 23\% [23\%] & 25\% [25\%] \\
    \(-\sfrac{5}{3}\) & 43\% [49\%] & 44\% [50\%] & 41\% [47\%] & 44\% [50\%]
    \\ \cline{1-5} \cline{1-5}
    \end{tabularx}
    \caption{
        \textbf{Extrapolated Small-scale Energy}
        Percentage of total energy integrating scales in the decade spanning 10--100 km.
        Values in brackets (\([\cdot]\)) arise from integrating all scales smaller than \(100~\)km assuming a constant power-law scaling of \(k^{-\alpha}\).
    }
    \label{table:sub100km_energy}
\end{table*}

\begin{figure}[ht]
    \centering
    \includegraphics[scale=1]{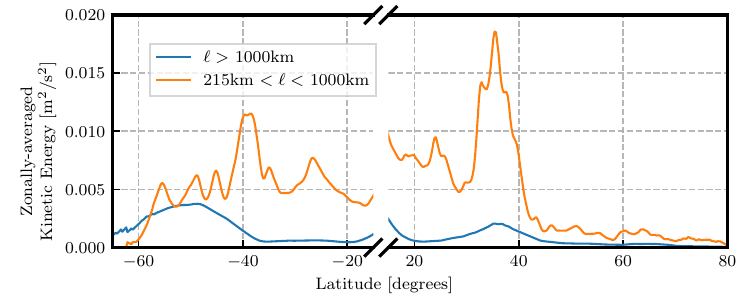}
    \caption{
        Time- and zonally-averaged kinetic energy computed from AVISO within selected length-scale bands (see in-set legend) as a function of latitude. We can see that the Antarctic Circumpolar Current has significant energy at scales ${>10^3~}$km, while the North has significant energy within ${\approx 30^\circ}$N-$40^\circ$N where the Western Boundary Currents are located. 
        Note that the latitude axis is broken to exclude the band [$15^\circ$S, $15^\circ$N].
    }
    \label{fig:zonally_avg_coarse_KE}
\end{figure}

\paragraph*{ Zonally-Averaged Coarse Energy }
In Figure~\ref{fig:zonally_avg_coarse_KE} we plot the zonally-averaged kinetic energy for selected length-scale bands. 
Scales larger than $10^3$~km (blue plot in Fig.~\ref{fig:zonally_avg_coarse_KE}) have a dominant contribution from latitudes [60$^\circ$S, 40$^\circ$S], which roughly corresponds with the ACC.
However, these latitudes are no longer dominant when considering the band of smaller scales: 215 km ${<\ell<}$ $10^3$~km. 
These scales (orange plot in Fig.~\ref{fig:zonally_avg_coarse_KE}) show a distinct signal at latitudes [30$^\circ$N, 40$^\circ$N], which roughly aligns with the Gulf Stream and Kuroshio.
There is also a weaker signal at latitudes [40$^\circ$S, 35$^\circ$S], which roughly aligns with the Agulhas and the Brazil-Malvinas currents.

\subsection{Spatio-temporal decomposition}
\label{sec:Spatio-temporalDecomposition}
In this section, we present results from coarse-graining in both space and time to reveal all the length-scales present in the time-averaged currents up to 9-year temporal mean. Our analysis demonstrates a way for comparing data from satellite analysis (AVISO) and numerical models (NEMO).

The approach consists of measuring the filtering spectrum of a temporally-smoothed version of the original velocity field. The latter is obtained from a running window time average, 
\begin{linenomath*}
\be
\langle \bu \rangle_\tau (\bx,t) = \frac{1}{\tau} \int_{t-\tau/2}^{t+\tau/2} \bu(\bx,t') \, \mathrm{d}t',
\label{eq:time-window-vel}
\ee
\end{linenomath*}
with $\tau$ the size of the time window. Note that a running window time-average in equation \eqref{eq:time-window-vel} is similar to spatial coarse-graining (equation~\eqref{def:filtering}) since 
\begin{linenomath*}
\be
\langle\langle F \rangle_\tau\rangle_\tau\ne \langle F \rangle_\tau. 
\ee
\end{linenomath*}
Combining equation~(\ref{eq:filt_spect}) with equation~(\ref{eq:time-window-vel}) allows us to   measure the filtering energy spectrum of the time-smoothed field
\begin{linenomath*}
\be
\olE(k_\ell,\tau)= 
\angp{\frac{d}{d k_\ell}\bracep{ \frac{1}{2}\absv{\langle\OL{\bu}_\ell\rangle_{\tau}}^2 } }
= 
\angp{ \frac{d}{d k_\ell} \bracep{ \ME_{\ell,\tau} } },
\label{eq:time-window-spect}
\ee
\end{linenomath*}
where we introduced
\begin{linenomath*}
\be
\ME_{\ell,\tau}(\bx,t)= \frac{1}{2} \absv{\angp{ \OL\bu_\ell}_\tau }^2,
\ee
\end{linenomath*}
which is the cumulative spectrum of the temporally-smoothed field. As indicated, $\ME_{\ell,\tau}(\bx,t)$ is a function of both the size of the time window, $\tau$, and the spatial kernel, $\ell$.

\begin{figure}[ht]
    \centering
    \includegraphics[scale=1]{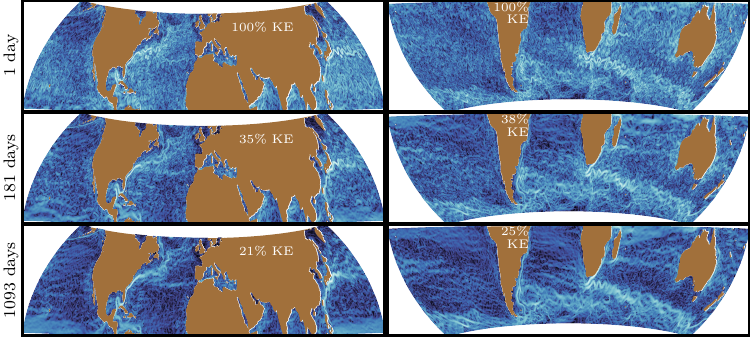}
    \caption{
        The surface geostrophic kinetic energy from the temporally coarse-grained flow, $\ME_{\ell=0,\tau}$, in the North (left column) and South (right column) from AVISO. 
        The top row shows the original 1-day averaged flow.
        The middle and bottom rows show the kinetic energy from the flow when averaged with a ${\approx 6~}$months time window and a ${\approx3~}$years time window, respectively, with the kinetic energy decreasing with an increasing time window. Each panel indicates the $\%$ of kinetic energy remaining relative to the 1-day top row.
    }
    \label{fig:time-windw-AVISO}
\end{figure}

\paragraph*{Time-Averaged Spatial Maps}
We show the time-smoothed energy map, $\ME_{\ell=0,\tau}$, in Figure~\ref{fig:time-windw-AVISO} from AVISO. Here, the two columns compare results from the North and the South regions, while different rows compare results with different time windows, $\tau$. From these maps we can see that increasing $\tau$ from one day to 1093 days reduces the energy down to ${\approx21\%}$ (${\approx25\%}$) of the original total energy in the North (South). Hence, averaging over three years brings the energy down to values comparable to those over the full nine years obtained in the previous section by the Reynolds averaging decomposition, where we found that $MKE$ accounts for  ${\approx20\%}$ of the total energy in the extra-tropics. This result indicates that temporal averaging converges quickly for the geostrophic kinetic energy, and using longer time records does not significantly alter the partitioning between the temporal mean and fluctuating components of the surface geostrophic ocean flow.

\subsection{Spatio-temporal comparison of AVISO and NEMO}
\label{sec:Spatio-temporal}

\begin{figure}[ht]
    \centering
    \includegraphics[scale = 1]{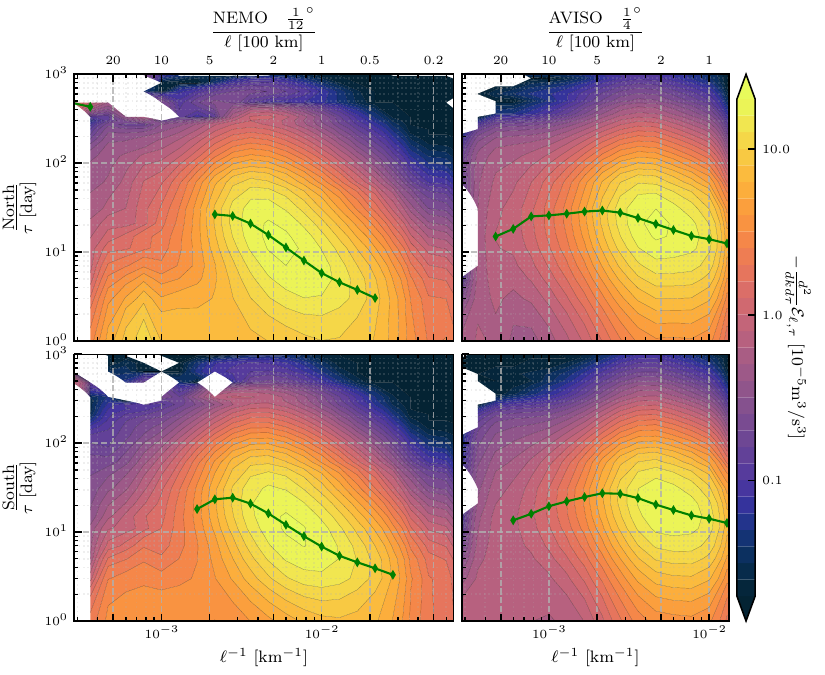}
    \caption{ 
        Combined spatio-temporal coarse-graining producing 2D spectra, ${-\partial_{\tau}\partial_{k_\ell} \ME_{\ell,\tau}}$ from \textbf{[left]} \(\sfrac{1}{12}^\circ\) NEMO and \textbf{[right]} AVISO, averaged over the \textbf{[top]} NH and \textbf{[bottom]} SH.
        Mesoscale energy predominantly peaks on length-scales of 100-200~km and time-scales of 1-3 weeks.
        Green diamonds indicate, for each \(\ell\), the \(\tau\) at which spectral power is maximized (c.~f. Fig~\ref{fig:peak_taus}).
    }
    \label{fig:spectra-tau-k}
\end{figure}

We now demonstrate using a spatio-temporal coarse-graining, which may complement current efforts to disentangle balanced from unbalanced motions in SSH-derived flows. 
Figure~\ref{fig:spectra-tau-k} presents space-time 2-D spectra, $-\langle \frac{d}{d \tau}\frac{d}{d k_\ell} \{ \ME_{\ell,\tau} \} \rangle$, which decomposes the energy as measured from AVISO and NEMO. 
In the left (right) column of Figure~\ref{fig:spectra-tau-k} we show the isolevels of space-time spectra from NEMO (AVISO).
Note that the NEMO spectra extend to smaller length scales due to having higher spatial resolution, but that the panels have consistent spacing / aspect ratios.
The most pronounced difference is that the AVISO isocontours are more circular, while NEMO isocontours or more elongated and tilted, hinting at an \({\ell-\tau}\) relationship.
In both datasets, energy peaks at approximately \(\ell=200~\)km and \(\tau=2-3~\)weeks.

\begin{figure}[ht]
    \centering
    \includegraphics[scale = 1]{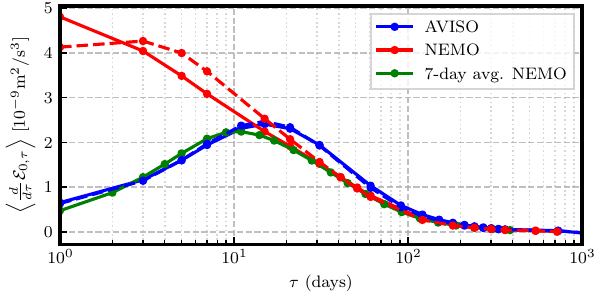}
    \caption{
        Evidence that the disagreement between AVISO and NEMO over time-scales ${\lesssim10~}$days is due to temporal averaging used in generating the gridded AVISO product. Here, we show temporal spectra from AVISO (blue) and NEMO (red) in the North (solid lines) and South (dashed lines), which disagree over ${\tau\lesssim10~}$days as in Figure~\ref{fig:spectra-tau-k}. However, the temporal spectra from NEMO agree with those from AVISO after applying a 7-day temporal smoothing to the original NEMO velocities (green). This result supports our hypothesis that AVISO is missing dynamical information at time-scales less than 10 days due to temporal smoothing over all length-scales.
    }
    \label{fig:repeat-tau}
\end{figure}

\paragraph*{Time-averaging to Align Spectra}
Remember that for the entire analysis in this paper, we are using 1-day averages of SSH to derive velocity from the NEMO data. While the SSH from AVISO is also available daily, it is effectively averaged over longer periods of time to produce gridded SSH maps from along-track altimeter data. 
We propose that the difference between isocontours from AVISO and NEMO in Figure~\ref{fig:spectra-tau-k} comes from the optimal interpolation used to produce the gridded AVISO product \cite{pujol2016}, which is necessary to construct the global maps from satellite altimeters' along-track data. 
To support this hypothesis, in Figure~\ref{fig:repeat-tau} we show the spectra as a function of $\tau$ measured from AVISO and NEMO. 
In this plot, we have repeated the analysis of the NEMO spectra after passing the data through a 7-day running time average (green line), which reproduces the time average over the satellite orbits. 
We can see that the green curve overlaps the AVISO measurement (blue) very closely, supporting our hypothesis. 
This is similar to what was done in \citeA{arbic2014geostrophic,khatri2018surface} who were comparing the cascade from AVISO and model data and determined that AVISO's spectral fluxes can be reproduced from model data after filtering the latter in both space and time.

\paragraph*{ Possible Role of Unbalanced Motions }
What component of the flow could be yielding the discrepancy between NEMO and AVISO? 
The most obvious possibility is unbalanced motion present in the 1-day mean SSH fields of NEMO that is absent from AVISO due to the effective weekly averaging required for gridding the satellite measurements. 
However, unbalanced motion had been believed to be important mostly over length-scales ${\lesssim 100}$~km and time-scales ${\lesssim2}$~days (e.g. \citeA{richman2012inferring,Qiuetal2018jpo}). 
If our conjecture is correct, it would imply that unbalanced motion is present at all scales between $200~$km to $10^3~$km, with significant differences even between ${1\text{-}2\times10^3~}$km and ${\tau\approx 1\text{-}10~}$days as shown in Fig.~\ref{fig:spectra-tau-k}, requiring averaging over a few days to be removed.  
Isolating balanced from unbalanced motions (e.g. \citeA{Buhleretal2014JFM}) is an active research topic that is beyond the scope of this work. 
Another possible explanation can be found in the time-smoothing of balanced motions, which is inherent in the construction of the AVISO dataset. Indeed in~\cite{Arbicetal13,arbic2014geostrophic} they removed high-frequency motions with a 3-day low-pass filter before applying spectral analysis and they obtained similar results as the ones we observed here.

\subsubsection{ Relating Time-scale to Length-scale }
\begin{figure}
    \centering
    \includegraphics[scale=1]{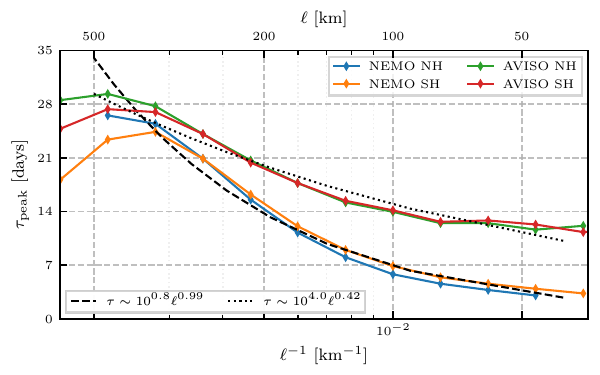}
    \caption{ 
        \textbf{ Mesoscale \(\tau\)-\(\ell\) Relationship }
        For each filter scale \((\ell)\), the time-scale \((\tau_{\mathrm{peak}})\) for which the power spectrum \({(-\partial_{\tau}\partial_{k_\ell} \ME_{\ell,\tau}})\) is maximized. While dashed lines show regression fits (see legend for regression formulas), which express \(\tau\) [s] in terms of \(\ell\) [m].
    }
    \label{fig:peak_taus}
\end{figure}

As discussed, Fig.~\ref{fig:spectra-tau-k} shows a clear mesoscale spectral structure centered roughly on 200~km and 14~days.
In Figure \ref{fig:peak_taus} we present for each spatial scale \(\ell\), the time-scale \(\tau\) for which \({-\partial_{\tau}\partial_{k_\ell} \ME_{\ell,\tau}}\) is maximized.
We use cubic interpolation in the \(\tau\)-dimension to compensate for only having data points for an odd integer number of days.
These results are broadly similar between hemispheres, however, there are noticeable disagreements between NEMO and AVISO.
The two agree on the time scale of the largest mesoscales (400--500km), with AVISO consistently yielding longer time scales than NEMO for smaller \(\ell\).
NEMO presents \({\tau\sim\ell}\) over the mesoscale band, while AVISO gives \({\tau\sim\ell^{0.4}}\).

\subsubsection{Connection to Space-Time Spectra in the Literature}

Figure~\ref{fig:spectra-tau-k} shows the importance of performing a combined spatio-temporal decomposition to access all information in the data. 
Our method is similar to frequency-wavenumber analysis performed within Fourier boxes by several recent studies:  \citeA{arbic2014geostrophic} were interested in mesoscale-driven intrinsic low-frequency variability, while \citeA{Savageetal2017jgr,Qiuetal2018jpo,Torresetal2018jgr} were primarily motivated by isolating the unbalanced motions from SSH-derived velocities. 
Our Figure~\ref{fig:spectra-tau-k} is analogous, for example, to Figure 4 in \citeA{arbic2014geostrophic} and to Figure 3 in \citeA{Torresetal2018jgr}, although they analyzed higher frequencies than those that are available in the datasets that we study here. It is important to stress that high-frequency forcing was not employed in the production of the NEMO model data used in our work and high-frequency motions are not our current focus of interest, while the latter works employed models with simultaneous atmospheric and tidal forcing which entails the formation of an internal gravity wave continuum spectrum as first described in~\cite{muller2015favors}.
However, as we mentioned in the introduction, the coarse-graining approach gives us access to the \emph{global} energy budget and, moreover, frees us from the limitations of Fourier boxes and the required tapering and detrending. 
As such, the approach here complements previous frequency-wavenumber analysis by allowing us to access much larger length-scales. 
 
A common feature between our Figure~\ref{fig:spectra-tau-k} and those in previous studies is a slight elongation of isocontours along the diagonal from  small to large spatio-temporal scales in the main panel of our Figure~\ref{fig:spectra-tau-k}. 
Such elongation is most prominent in Figure 3 of \citeA{Torresetal2018jgr}, who were probing scales ${<100}~$km and from roughly $3~$hours to $40~$days. 
The diagonal elongation of isocontours represents a slight tendency for larger length-scales to have longer time-scales.

However, we emphasize that unlike in \citeA{Torresetal2018jgr}, such tendency is only \emph{slight} over the larger scales we analyze here. 
In fact, an important take-away from Figure~\ref{fig:spectra-tau-k} is that all length-scales evolve over a wide range of time-scales. Consider, for example, ${\ell\approx 500~}$km in the left column of Figure~\ref{fig:spectra-tau-k} at different $\tau$ values. 
We see that the isoline is almost vertical over ${\tau\approx 5~}$days to ${\tau\approx 50~}$days, indicating that flow at $500~$km has an equal contribution from all these time-scales. 
We also see that both AVISO and NEMO isolines get flatter (stretched horizontally) as $\tau$ increases, such that at ${\tau\approx300~}$days, there is almost equal energy at all scales between ${\approx100~}$km and ${\approx10^3~}$km.

\section{Conclusions}
\label{sect:Conclusions}

\subsection{Summary of the main results}
In this paper we expanded on a recent calculation of the first global energy spectrum of the ocean's surface geostrophic circulation \cite{Storer2022} using the coarse-graining (CG) method. Our analysis here gives new insights into the oceanic circulation. The method is implemented in an open-source software, FlowSieve, that can be accessed at \url{https://github.com/husseinaluie/FlowSieve}. 

In this work, we compare quantitatively the CG and the spherical harmonics decompositions. While the two methods yield qualitatively consistent domain-averaged results, spherical harmonics spectra are too noisy at gyre scales. More importantly, spherical harmonics are inherently global and cannot provide local information connecting scales with currents geographically.

Similarly, we have estimated that the RMS velocity of the mesoscales is globally around $15$cm/s, but it increases up to 30--40~cm/s in the Kuroshio or the Gulf Stream and up to 16--28~cm/s in the ACC. We find notable hemispheric asymmetry in mesoscale energy-per-area, which is higher in the north due to continental boundaries.

In this paper, we applied the coarse-graining approach to the Reynolds decomposed fields, namely the time-mean and the time-varying terms of the ocean surface currents. Results in this direction highlight that while the time-varying term is largely dominated by the mesoscales, ($\sim 98\%$ of total energy), the time-mean component also has a majority (up to 70\%) contribution from the mesoscale circulation. This highlights the preponderance of `standing' small-scale structures in the global ocean. It also shows that Reynolds decomposition is an effective method for disentangling eddy structures from the flow.

 By coarse-graining in both space and time, we have shown that every length-scale evolves over a wide range of time-scales. This result makes us appreciate the significance of temporally coherent (even stationary) forcing mechanisms acting on the mesoscales, such as bottom topography and continental boundaries.
An important new contribution of this work is the spatio-temporal spectra of the geostrophic currents.
These 2D spectra highlight how the mesoscales while peaking at \({\approx(200~\mathrm{km},2~\mathrm{weeks})}\), are not only diffused over a range of spatial scales, but also vary over a wide range of temporal scales.
Further, we extract the dominant time-scale, \(\tau_{\mathrm{peak}}\) for each filter scale in the mesoscale band, and find that NEMO predicts \({\tau\sim\ell}\), which leads to a length scale-independent advective velocity of 0.15--0.2~cm/s.
In contrast, AVISO demonstrates consistently longer dominant time-scales, and a shallower relationship of \({\tau\sim\ell^{0.4}}\), both of which are likely results of the time averaging needed to extract the AVISO velocity maps.

\subsection{Coarse-graining and the filtering spectrum}

The coupling between different length- and time-scales and between different geographic regions presents a major difficulty in understanding, modeling, and predicting oceanic circulation and mixing. 
Indeed, the oceanic kinetic energy budget is estimated to suffer from large uncertainties \cite{FerrariWunsch09}. 
A major reason behind these difficulties is a lack of scale-analysis methods that are appropriate in the \emph{global} ocean.
In this paper, we have demonstrated the versatility of coarse-graining in serving as a robust scale-analysis method for the global ocean circulation that complements existing methods. 
The approach is very general, allows for probing the dynamics simultaneously in scale and in space, and is not restricted by assumptions of homogeneity or isotropy commonly required for traditional methods such as Fourier or structure-function analysis. Coarse-graining  includes wavelet analysis as a special case with the proper choice of convolution kernel \cite{sadek2018extracting}. 
Coarse-graining offers a way to probe and quantify the energy budget at different length-scales globally while maintaining local information about the heterogeneous oceanic regions. 
We view this work as an important step toward constructing a scale-aware global Lorenz Energy Cycle for the ocean circulation \cite{loose2023diagnosing}.

\appendix

\section{Deforming the kernel around land}
\label{appendix:deforming_kernel}

As outlined in section \ref{sect:meth-CG}, filtering with a constant kernel while treating land as zero-velocity water and including land cells (``Fixed Kernel w/ Land'') in the final tally is guaranteed to conserve $100\%$ of the energy, while excluding land cells and integrating only over water cells (``Fixed Kernel w/o Land'') leads to a loss of about 11\% of the total kinetic energy at a filter scale of $2,000$~km (see Figure~\ref{fig:energy-leak}).
This result follows since some of the kinetic energy `smears' onto the land cells, which are then excluded from the spatial integrals.

An alternative approach is to deform the kernel around land (``Deforming Kernel'') so that only water cells are incorporated in the filtering operation.
This approach has the advantage of not needing to treat land as water, yet we have shown in Figure~\ref{fig:energy-leak} that this choice still does not conserve $100\%$ of the energy, sometimes even yielding larger values, albeit still within 1\% error. Here, we explain why a deforming a kernel cannot be expected to yield $100\%$ of the energy, unlike the ``Fixed Kernel w/ Land.''

To illustrate how the loss of energy conservation can happen with the Deforming Kernel method, consider a one-dimensional domain with five equally spaced points and a simple kernel that has a weight of 2 at the target point, 1 at neighbouring points, and 0 otherwise.

If the domain were periodic then the filtering operation could be represented as the matrix
\[
G:=
\left[
\begin{array}{ccccc}
    \sfrac{1}{2} & \sfrac{1}{4} & 0            & 0            & \sfrac{1}{4}  \\
    \sfrac{1}{4} & \sfrac{1}{2} & \sfrac{1}{4} & 0            & 0 \\
    0            & \sfrac{1}{4} & \sfrac{1}{2} & \sfrac{1}{4} & 0 \\
    0            & 0            & \sfrac{1}{4} & \sfrac{1}{2} & \sfrac{1}{4}  \\
    \sfrac{1}{4} & 0            & 0            & \sfrac{1}{4} & \sfrac{1}{2}
\end{array}
\right]
\]
such that \({\overline{\mathrm{KE}} = G\cdot\mathrm{KE}}\), where \(\mathrm{KE}\) is a column vector.
Note that the sum of each row of \(G\) is 1, a result of normalizing the kernel (assuming a grid spacing of 1 for simplicity).
Domain integrating in this scenario is simply left-multiplying by the row vector \({S:=[1,1,1,1,1]}\), which is equivalent to taking a column-wise sum.
Since \({S\cdot G = S}\), \( {S\cdot\overline{\mathrm{KE}} = S\cdot G\cdot\mathrm{KE} = S\cdot\mathrm{KE}}\),
and so the domain-integrated kinetic energy is conserved.

However, if the domain is non-periodic (such as if the edges were `land'), then the deforming kernel that excludes anything outside the boundaries would be
\[
G:=
\left[
\begin{array}{ccccc}
    \sfrac{2}{3} & \sfrac{1}{3} & 0            & 0            & 0  \\
    \sfrac{1}{4} & \sfrac{1}{2} & \sfrac{1}{4} & 0            & 0 \\
    0            & \sfrac{1}{4} & \sfrac{1}{2} & \sfrac{1}{4} & 0 \\
    0            & 0            & \sfrac{1}{4} & \sfrac{1}{2} & \sfrac{1}{4}  \\
    0            & 0            & 0            & \sfrac{1}{3} & \sfrac{2}{3}
\end{array}
\right]
\]
In this case, \(S\cdot G = [ \sfrac{11}{12}, \sfrac{13}{12}, 1, \sfrac{13}{12}, \sfrac{11}{13} ] \neq S\), and so in general \(S\cdot\overline{\mathrm{KE}} \neq S\cdot\mathrm{KE}\).
Moreover, there is no guarantee that \(S\cdot\overline{\mathrm{KE}} \leq S\cdot\mathrm{KE}\), and so it may be that the total filtered kinetic energy exceeds the total unfiltered kinetic energy.

As observed, in general, the error arising from deforming the kernel will be much smaller than that of treating land as zero-velocity water and only integrating over true water cells, especially for large filter kernels. However, again, it is worth recognizing that deforming the kernel does not guarantee energy conservation.
To fully conserve energy and maintain commutativity with differentiation, we choose the ``Fixed Kernel w/ Land'' option, which treats land as zero-velocity water and includes land cells in spatial integrals to compute total energy. 

\section{Reynolds averaging spectra on AVISO dataset}
\label{appendix:ra_aviso}

\begin{figure}[h!]
    \centering
    \includegraphics[scale=1]{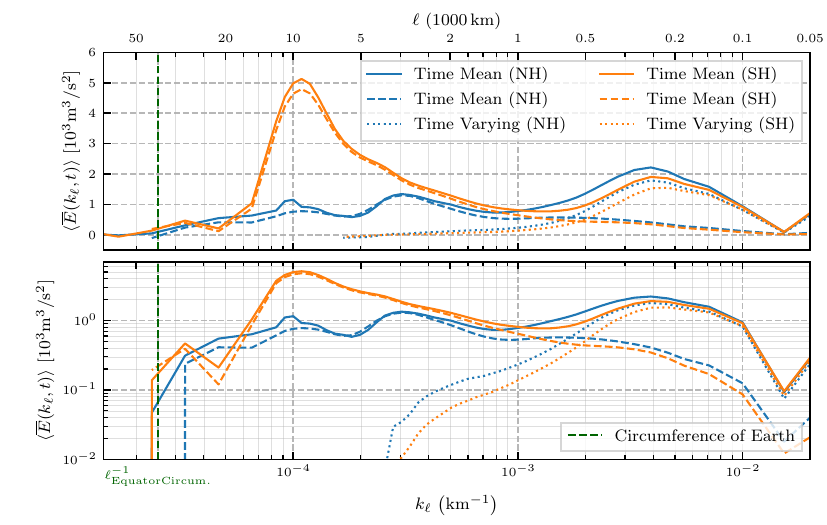}
    \caption{ 
        \textbf{Power Spectra}
        Filtering spectra obtained following eq.~\eqref{eq:filt_spect} for the full (solid lines), time mean (dashed times), and time-varying (dotted liens) ssh-derived geostrophic velocity from the AVISO dataset.
        Note that both top and bottom panels show the same data, but using lin-log and log-log scales respectively.
    }
    \label{fig:ra_spectra_aviso}
\end{figure}

Fig~\ref{fig:ra_spectra_aviso} reports the energy spectra for the time-mean and time-varying Reynolds averaging components obtained from the 9-year AVISO dataset. Results are in very good agreement with the spectra obtained from NEMO dataset, presented in Fig.~\ref{fig:spectra}. 
The values obtained from the two datasets are nearly identical, with the AVISO dataset having less small-scale energy owing to having a lower resolution.

\section{ Geographic Definitions for Current Regions }
\label{appendix:region_defs}

\begin{figure}[ht]
    \centering
    \includegraphics{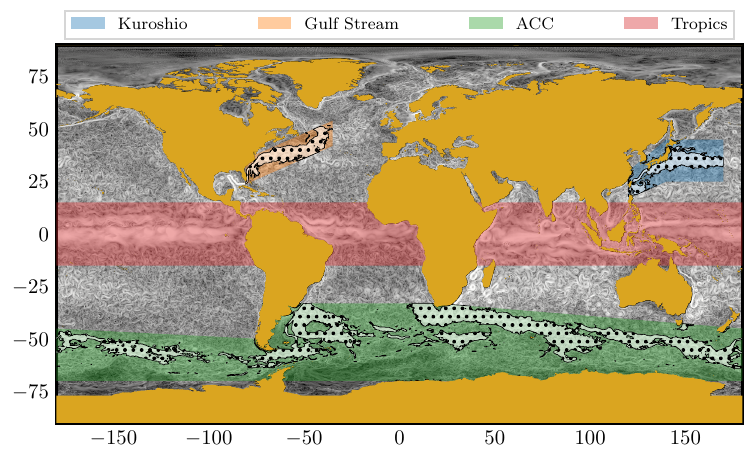}
    \caption{ 
        Illustration of the geographic region definitions (equations \eqref{eq:reg_def:begin}--\eqref{eq:reg_def:end}), plotted over a sample velocity field for reference.
        Note that `North of Tropics' and `South of Tropics' are not included, but are simply the portions North and South of `Tropics'.
        For `Kuroshio', `Gulf Stream', and `ACC', the smaller contoured region with dots shows the region definition with an additional KE mask.
    }
    \label{fig:region_masks}
\end{figure}

Equations~\eqref{eq:reg_def:begin}--\eqref{eq:reg_def:end} outline the geographic constraints used to define the various regions used in Table~\ref{table:rms_vels}.
In each definition, \(\lambda\) is longitude in degrees, ranging from \(-180\) to 180, and \(\phi\) is latitude in degrees, ranging from \(-90\) to 90.
Additionally, any overlap with land is removed from the region definition, so that only water cells are included.
The region masks are presented in Figure~\ref{fig:region_masks}. 

\paragraph*{Energy Masking}
Following \citeA{Raietal2021}, subsets of the regions defined in equations~\eqref{eq:reg_def:begin}--\eqref{eq:reg_def:end} are produced by further restricting to areas with sufficiently high ``masking KE''.
For these purposes, a combination of time-mean and time-varying KE is used such that 
\begin{equation}
    \text{Masking KE} = \frac{1}{2}\rho_0\angp{\mathbf{u}}^2 + \frac{1}{2}\rho_0\angp{\paren{\mathbf{u}-\angp{\mathbf{u}}}^2}.
\end{equation}
Taking \(\rho_0=1025\), a cut-off of \(\text{Masking KE}>50\) is applied to the Gulf Stream and Kuroshio, and \(\text{Masking KE}>30\) to the ACC.
The KE-masked regions are illustrated with dots in Figure~\ref{fig:region_masks}.

\begin{align}
\textbf{North of Tropics :}
&\quad
\phi > 15^\circ
\label{eq:reg_def:begin}\\
\textbf{Kuroshio :}
&\quad
\bracep{ 120^\circ < \lambda < 170^\circ } 
\notag\\
&\quad
\texttt{ and } 
\bracep{ 17^\circ < \phi < 45^\circ } 
\notag\\
&\quad
\texttt{ and }
\bracep{\phi \leq (\sfrac{3}{4})\lambda - 60^\circ }
\notag\\
&\quad
\texttt{ and }
\bracep{\texttt{not}\paren{ \phi<25^\circ \texttt{ and } \lambda\geq140^\circ }}
\notag\\
&\quad
\texttt{ and }
\bracep{\texttt{not}\paren{ \lambda\leq140^\circ \texttt{ and } \phi<(\sfrac{2}{5})\lambda-31^\circ }}
\\
\textbf{Gulf Stream :}
&\quad
\bracep{ -80.75^\circ < \lambda < -35^\circ }
\texttt{ and }
\bracep{ \absv{\phi - (\sfrac{2}{5})\lambda - 62^\circ} \leq 6^\circ }
\\
\textbf{South of Tropics :}
&\quad
\phi < -15^\circ
\\
\textbf{ACC :} 
&\quad
\bracep{ -70^\circ < \phi < -33^\circ }
\notag\\
&\quad
\texttt{ and }
\bracep{ \texttt{not}\paren{ \lambda < -72^\circ } \texttt{ and } \phi>-(\sfrac{5}{108})\lambda-\sfrac{160}{3}^\circ }
\notag\\
&\quad
\texttt{ and }
\bracep{ \texttt{not}\paren{ \lambda > 20^\circ } \texttt{ and } \phi>-(\sfrac{3}{40})\lambda-\sfrac{63}{2}^\circ }
\label{eq:reg_def:end}
\end{align}

\section*{Open Research}

This study has been conducted using E.U. Copernicus Marine Service Information. The product identifier of the AVISO dataset used in this work is ``\sloppy{SEALEVEL\_GLO\_PHY\_L4\_REP\_OBSERVATIONS\_008\_047}'', and can be downloaded at \url{https://marine.copernicus.eu/services-portfolio/access-to-products/}. The product identifier of the NEMO dataset is ``\sloppy{GLOBAL\_ANALYSISFORECAST\_PHY\_CPL\_001\_015}'', and is available at \url{https://marine.copernicus.eu/services-portfolio/access-to-products/}.  The source code for the coarse-graining software can be freely downloaded from https://github.com/husseinaluie/FlowSieve .

\acknowledgments
We thank B. Reichl, D. Balwada, M. Jansen, and S. Rai for their valuable discussions and comments. 
This research was funded by US NASA grant 80NSSC18K0772 and NSF grant OCE-2123496. HA was also supported by US DOE grants DE-SC0014318, DE-SC0020229, DE-SC0019329, NSF grant PHY-2020249, and US NNSA grants DE-NA0003856, DE-NA0003914. Computing time was provided by NERSC under Contract No. DE-AC02-05CH11231 and NASA's HEC Program through NCCS at Goddard Space Flight Center. MB was also supported by the European Research Council (ERC) under the European Union’s Horizon 2020 research and innovation programme (Grant Agreement No. 882340). HK was supported by UK NERC grant NE/T013494/1.

%% ------------------------------------------------------------------------ %%
%% References and Citations
%%%%%%%%%%%%%%%%%%%%%%%%%%%%%%%%%%%%%%%%%%%%%%%

%\bibliography{biblio}

\begin{thebibliography}{}

\end{thebibliography}


\begin{thebibliography}{}

\bibitem [\protect \citeauthoryear {%
Aluie%
}{%
Aluie%
}{%
{\protect \APACyear {2013}}%
}]{%
Aluie13}
\APACinsertmetastar {%
Aluie13}%
\begin{APACrefauthors}%
Aluie, H.%
\end{APACrefauthors}%
\unskip\
\newblock
\APACrefYearMonthDay{2013}{{\APACmonth{03}}}{}.
\newblock
{\BBOQ}\APACrefatitle {{Scale decomposition in compressible turbulence}}
  {{Scale decomposition in compressible turbulence}}.{\BBCQ}
\newblock
\APACjournalVolNumPages{Physica D: Nonlinear Phenomena}{247}{1}{54--65}.
\PrintBackRefs{\CurrentBib}

\bibitem [\protect \citeauthoryear {%
Aluie%
}{%
Aluie%
}{%
{\protect \APACyear {2017}}%
}]{%
Aluie17}
\APACinsertmetastar {%
Aluie17}%
\begin{APACrefauthors}%
Aluie, H.%
\end{APACrefauthors}%
\unskip\
\newblock
\APACrefYearMonthDay{2017}{{\APACmonth{01}}}{}.
\newblock
{\BBOQ}\APACrefatitle {{Coarse-grained incompressible magnetohydrodynamics:
  analyzing the turbulent cascades}} {{Coarse-grained incompressible
  magnetohydrodynamics: analyzing the turbulent cascades}}.{\BBCQ}
\newblock
\APACjournalVolNumPages{New Journal of Physics}{19}{}{025008}.
\PrintBackRefs{\CurrentBib}

\bibitem [\protect \citeauthoryear {%
Aluie%
}{%
Aluie%
}{%
{\protect \APACyear {2019}}%
}]{%
aluie2019convolutions}
\APACinsertmetastar {%
aluie2019convolutions}%
\begin{APACrefauthors}%
Aluie, H.%
\end{APACrefauthors}%
\unskip\
\newblock
\APACrefYearMonthDay{2019}{}{}.
\newblock
{\BBOQ}\APACrefatitle {Convolutions on the sphere: commutation with
  differential operators} {Convolutions on the sphere: commutation with
  differential operators}.{\BBCQ}
\newblock
\APACjournalVolNumPages{GEM-International Journal on Geomathematics}{10}{1}{9}.
\PrintBackRefs{\CurrentBib}

\bibitem [\protect \citeauthoryear {%
{Aluie}%
\ \BBA {} {Eyink}%
}{%
{Aluie}%
\ \BBA {} {Eyink}%
}{%
{\protect \APACyear {2009}}%
}]{%
AluieEyink09}
\APACinsertmetastar {%
AluieEyink09}%
\begin{APACrefauthors}%
{Aluie}, H.%
\BCBT {}\ \BBA {} {Eyink}, G.%
\end{APACrefauthors}%
\unskip\
\newblock
\APACrefYearMonthDay{2009}{{\APACmonth{11}}}{}.
\newblock
{\BBOQ}\APACrefatitle {{Localness of energy cascade in hydrodynamic turbulence.
  II. Sharp spectral filter}} {{Localness of energy cascade in hydrodynamic
  turbulence. II. Sharp spectral filter}}.{\BBCQ}
\newblock
\APACjournalVolNumPages{Phys. Fluids}{21}{11}{115108}.
\PrintBackRefs{\CurrentBib}

\bibitem [\protect \citeauthoryear {%
Aluie%
, Hecht%
\BCBL {}\ \BBA {} Vallis%
}{%
Aluie%
\ \protect \BOthers {.}}{%
{\protect \APACyear {2018}}%
}]{%
aluie2018mapping}
\APACinsertmetastar {%
aluie2018mapping}%
\begin{APACrefauthors}%
Aluie, H.%
, Hecht, M.%
\BCBL {}\ \BBA {} Vallis, G\BPBI K.%
\end{APACrefauthors}%
\unskip\
\newblock
\APACrefYearMonthDay{2018}{}{}.
\newblock
{\BBOQ}\APACrefatitle {Mapping the energy cascade in the North Atlantic Ocean:
  The coarse-graining approach} {Mapping the energy cascade in the north
  atlantic ocean: The coarse-graining approach}.{\BBCQ}
\newblock
\APACjournalVolNumPages{Journal of Physical Oceanography}{48}{2}{225--244}.
\PrintBackRefs{\CurrentBib}

\bibitem [\protect \citeauthoryear {%
{Aluie}%
, {Li}%
\BCBL {}\ \BBA {} {Li}%
}{%
{Aluie}%
\ \protect \BOthers {.}}{%
{\protect \APACyear {2012}}%
}]{%
Aluieetal12}
\APACinsertmetastar {%
Aluieetal12}%
\begin{APACrefauthors}%
{Aluie}, H.%
, {Li}, S.%
\BCBL {}\ \BBA {} {Li}, H.%
\end{APACrefauthors}%
\unskip\
\newblock
\APACrefYearMonthDay{2012}{{\APACmonth{06}}}{}.
\newblock
{\BBOQ}\APACrefatitle {{Conservative Cascade of Kinetic Energy in Compressible
  Turbulence}} {{Conservative Cascade of Kinetic Energy in Compressible
  Turbulence}}.{\BBCQ}
\newblock
\APACjournalVolNumPages{Astrophys. J. Lett.}{751}{}{L29}.
\PrintBackRefs{\CurrentBib}

\bibitem [\protect \citeauthoryear {%
Aluie%
\ \BBA {} Teeraratkul%
}{%
Aluie%
\ \BBA {} Teeraratkul%
}{%
{\protect \APACyear {2023}}%
}]{%
AluieTeeraratkul2023}
\APACinsertmetastar {%
AluieTeeraratkul2023}%
\begin{APACrefauthors}%
Aluie, H.%
\BCBT {}\ \BBA {} Teeraratkul, C.%
\end{APACrefauthors}%
\unskip\
\newblock
\APACrefYearMonthDay{2023}{}{}.
\newblock
{\BBOQ}\APACrefatitle {Theory of Large Eddy Simulation on the Sphere} {Theory
  of large eddy simulation on the sphere}.{\BBCQ}
\newblock
\APACjournalVolNumPages{}{in preparation}{}{}.
\PrintBackRefs{\CurrentBib}

\bibitem [\protect \citeauthoryear {%
Amores%
, Jord{\`a}%
, Arsouze%
\BCBL {}\ \BBA {} Le~Sommer%
}{%
Amores%
\ \protect \BOthers {.}}{%
{\protect \APACyear {2018}}%
}]{%
amores2018up}
\APACinsertmetastar {%
amores2018up}%
\begin{APACrefauthors}%
Amores, A.%
, Jord{\`a}, G.%
, Arsouze, T.%
\BCBL {}\ \BBA {} Le~Sommer, J.%
\end{APACrefauthors}%
\unskip\
\newblock
\APACrefYearMonthDay{2018}{}{}.
\newblock
{\BBOQ}\APACrefatitle {Up to what extent can we characterize ocean eddies using
  present-day gridded altimetric products?} {Up to what extent can we
  characterize ocean eddies using present-day gridded altimetric
  products?}{\BBCQ}
\newblock
\APACjournalVolNumPages{Journal of Geophysical Research:
  Oceans}{123}{10}{7220--7236}.
\PrintBackRefs{\CurrentBib}

\bibitem [\protect \citeauthoryear {%
Arbic%
\ \protect \BOthers {.}}{%
Arbic%
\ \protect \BOthers {.}}{%
{\protect \APACyear {2014}}%
}]{%
arbic2014geostrophic}
\APACinsertmetastar {%
arbic2014geostrophic}%
\begin{APACrefauthors}%
Arbic, B\BPBI K.%
, M{\"u}ller, M.%
, Richman, J\BPBI G.%
, Shriver, J\BPBI F.%
, Morten, A\BPBI J.%
, Scott, R\BPBI B.%
\BDBL {}Penduff, T.%
\end{APACrefauthors}%
\unskip\
\newblock
\APACrefYearMonthDay{2014}{}{}.
\newblock
{\BBOQ}\APACrefatitle {Geostrophic turbulence in the frequency--wavenumber
  domain: Eddy-driven low-frequency variability} {Geostrophic turbulence in the
  frequency--wavenumber domain: Eddy-driven low-frequency variability}.{\BBCQ}
\newblock
\APACjournalVolNumPages{Journal of Physical Oceanography}{44}{8}{2050--2069}.
\PrintBackRefs{\CurrentBib}

\bibitem [\protect \citeauthoryear {%
Arbic%
, Polzin%
, Scott%
, Richman%
\BCBL {}\ \BBA {} Shriver%
}{%
Arbic%
\ \protect \BOthers {.}}{%
{\protect \APACyear {2013}}%
}]{%
Arbicetal13}
\APACinsertmetastar {%
Arbicetal13}%
\begin{APACrefauthors}%
Arbic, B\BPBI K.%
, Polzin, K\BPBI L.%
, Scott, R\BPBI B.%
, Richman, J\BPBI G.%
\BCBL {}\ \BBA {} Shriver, J\BPBI F.%
\end{APACrefauthors}%
\unskip\
\newblock
\APACrefYearMonthDay{2013}{{\APACmonth{02}}}{}.
\newblock
{\BBOQ}\APACrefatitle {{On Eddy Viscosity, Energy Cascades, and the Horizontal
  Resolution of Gridded Satellite Altimeter Products*}} {{On Eddy Viscosity,
  Energy Cascades, and the Horizontal Resolution of Gridded Satellite Altimeter
  Products*}}.{\BBCQ}
\newblock
\APACjournalVolNumPages{Journal of Physical Oceanography}{43}{2}{283--300}.
\PrintBackRefs{\CurrentBib}

\bibitem [\protect \citeauthoryear {%
Arbic%
\ \protect \BOthers {.}}{%
Arbic%
\ \protect \BOthers {.}}{%
{\protect \APACyear {2012}}%
}]{%
arbic2012nonlinear}
\APACinsertmetastar {%
arbic2012nonlinear}%
\begin{APACrefauthors}%
Arbic, B\BPBI K.%
, Scott, R\BPBI B.%
, Flierl, G\BPBI R.%
, Morten, A\BPBI J.%
, Richman, J\BPBI G.%
\BCBL {}\ \BBA {} Shriver, J\BPBI F.%
\end{APACrefauthors}%
\unskip\
\newblock
\APACrefYearMonthDay{2012}{}{}.
\newblock
{\BBOQ}\APACrefatitle {Nonlinear cascades of surface oceanic geostrophic
  kinetic energy in the frequency domain} {Nonlinear cascades of surface
  oceanic geostrophic kinetic energy in the frequency domain}.{\BBCQ}
\newblock
\APACjournalVolNumPages{Journal of Physical Oceanography}{42}{9}{1577--1600}.
\PrintBackRefs{\CurrentBib}

\bibitem [\protect \citeauthoryear {%
Ballarotta%
\ \protect \BOthers {.}}{%
Ballarotta%
\ \protect \BOthers {.}}{%
{\protect \APACyear {2019}}%
}]{%
ballarotta2019resolutions}
\APACinsertmetastar {%
ballarotta2019resolutions}%
\begin{APACrefauthors}%
Ballarotta, M.%
, Ubelmann, C.%
, Pujol, M\BHBI I.%
, Taburet, G.%
, Fournier, F.%
, Legeais, J\BHBI F.%
\BDBL {}others%
\end{APACrefauthors}%
\unskip\
\newblock
\APACrefYearMonthDay{2019}{}{}.
\newblock
{\BBOQ}\APACrefatitle {On the resolutions of ocean altimetry maps} {On the
  resolutions of ocean altimetry maps}.{\BBCQ}
\newblock
\APACjournalVolNumPages{Ocean Science}{15}{4}{1091--1109}.
\PrintBackRefs{\CurrentBib}

\bibitem [\protect \citeauthoryear {%
Barkan%
\ \protect \BOthers {.}}{%
Barkan%
\ \protect \BOthers {.}}{%
{\protect \APACyear {2021}}%
}]{%
barkan2021oceanic}
\APACinsertmetastar {%
barkan2021oceanic}%
\begin{APACrefauthors}%
Barkan, R.%
, Srinivasan, K.%
, Yang, L.%
, McWilliams, J\BPBI C.%
, Gula, J.%
\BCBL {}\ \BBA {} Vic, C.%
\end{APACrefauthors}%
\unskip\
\newblock
\APACrefYearMonthDay{2021}{}{}.
\newblock
{\BBOQ}\APACrefatitle {Oceanic mesoscale eddy depletion catalyzed by internal
  waves} {Oceanic mesoscale eddy depletion catalyzed by internal waves}.{\BBCQ}
\newblock
\APACjournalVolNumPages{TBD}{submitted}{}{}.
\newblock
\begin{APACrefDOI} \doi{10.1002/essoar.10507068.1} \end{APACrefDOI}
\PrintBackRefs{\CurrentBib}

\bibitem [\protect \citeauthoryear {%
Biferale%
, Bonaccorso%
, Buzzicotti%
\BCBL {}\ \BBA {} Iyer%
}{%
Biferale%
\ \protect \BOthers {.}}{%
{\protect \APACyear {2019}}%
}]{%
Biferale2019SelfSim}
\APACinsertmetastar {%
Biferale2019SelfSim}%
\begin{APACrefauthors}%
Biferale, L.%
, Bonaccorso, F.%
, Buzzicotti, M.%
\BCBL {}\ \BBA {} Iyer, K\BPBI P.%
\end{APACrefauthors}%
\unskip\
\newblock
\APACrefYearMonthDay{2019}{Jul}{}.
\newblock
{\BBOQ}\APACrefatitle {Self-Similar Subgrid-Scale Models for Inertial Range
  Turbulence and Accurate Measurements of Intermittency} {Self-similar
  subgrid-scale models for inertial range turbulence and accurate measurements
  of intermittency}.{\BBCQ}
\newblock
\APACjournalVolNumPages{Phys. Rev. Lett.}{123}{}{014503}.
\newblock
\begin{APACrefDOI} \doi{10.1103/PhysRevLett.123.014503} \end{APACrefDOI}
\PrintBackRefs{\CurrentBib}

\bibitem [\protect \citeauthoryear {%
Bryan%
, Gent%
\BCBL {}\ \BBA {} Tomas%
}{%
Bryan%
\ \protect \BOthers {.}}{%
{\protect \APACyear {2014}}%
}]{%
Bryanetal14}
\APACinsertmetastar {%
Bryanetal14}%
\begin{APACrefauthors}%
Bryan, F\BPBI O.%
, Gent, P\BPBI R.%
\BCBL {}\ \BBA {} Tomas, R.%
\end{APACrefauthors}%
\unskip\
\newblock
\APACrefYearMonthDay{2014}{{\APACmonth{01}}}{}.
\newblock
{\BBOQ}\APACrefatitle {{Can Southern Ocean Eddy Effects Be Parameterized in
  Climate Models? }} {{Can Southern Ocean Eddy Effects Be Parameterized in
  Climate Models? }}.{\BBCQ}
\newblock
\APACjournalVolNumPages{Journal of Climate}{27}{1}{411--425}.
\PrintBackRefs{\CurrentBib}

\bibitem [\protect \citeauthoryear {%
B{\"u}hler%
, Callies%
\BCBL {}\ \BBA {} Ferrari%
}{%
B{\"u}hler%
\ \protect \BOthers {.}}{%
{\protect \APACyear {2014}}%
}]{%
Buhleretal2014JFM}
\APACinsertmetastar {%
Buhleretal2014JFM}%
\begin{APACrefauthors}%
B{\"u}hler, O.%
, Callies, J.%
\BCBL {}\ \BBA {} Ferrari, R.%
\end{APACrefauthors}%
\unskip\
\newblock
\APACrefYearMonthDay{2014}{{\APACmonth{10}}}{}.
\newblock
{\BBOQ}\APACrefatitle {{Wave-vortex decomposition of one-dimensional ship-track
  data}} {{Wave-vortex decomposition of one-dimensional ship-track
  data}}.{\BBCQ}
\newblock
\APACjournalVolNumPages{Journal of Fluid Mechanics}{756}{}{1007--1026}.
\PrintBackRefs{\CurrentBib}

\bibitem [\protect \citeauthoryear {%
Busecke%
\ \BBA {} Abernathey%
}{%
Busecke%
\ \BBA {} Abernathey%
}{%
{\protect \APACyear {2019}}%
}]{%
BuseckeAbernathey2019}
\APACinsertmetastar {%
BuseckeAbernathey2019}%
\begin{APACrefauthors}%
Busecke, J\BPBI J\BPBI M.%
\BCBT {}\ \BBA {} Abernathey, R\BPBI P.%
\end{APACrefauthors}%
\unskip\
\newblock
\APACrefYearMonthDay{2019}{{\APACmonth{01}}}{}.
\newblock
{\BBOQ}\APACrefatitle {{Ocean mesoscale mixing linked to climate variability}}
  {{Ocean mesoscale mixing linked to climate variability}}.{\BBCQ}
\newblock
\APACjournalVolNumPages{Science Advances}{5}{}{eaav5014--}.
\PrintBackRefs{\CurrentBib}

\bibitem [\protect \citeauthoryear {%
{Buzzicotti}%
, {Aluie}%
, Biferale%
\BCBL {}\ \BBA {} Linkmann%
}{%
{Buzzicotti}%
\ \protect \BOthers {.}}{%
{\protect \APACyear {2018}}%
}]{%
buzzicotti2018energy}
\APACinsertmetastar {%
buzzicotti2018energy}%
\begin{APACrefauthors}%
{Buzzicotti}, M.%
, {Aluie}, H.%
, Biferale, L.%
\BCBL {}\ \BBA {} Linkmann, M.%
\end{APACrefauthors}%
\unskip\
\newblock
\APACrefYearMonthDay{2018}{}{}.
\newblock
{\BBOQ}\APACrefatitle {Energy transfer in turbulence under rotation} {Energy
  transfer in turbulence under rotation}.{\BBCQ}
\newblock
\APACjournalVolNumPages{Physical Review Fluids}{3}{3}{034802}.
\PrintBackRefs{\CurrentBib}

\bibitem [\protect \citeauthoryear {%
Buzzicotti%
\ \BBA {} Clark Di~Leoni%
}{%
Buzzicotti%
\ \BBA {} Clark Di~Leoni%
}{%
{\protect \APACyear {2020}}%
}]{%
buzzicotti2020synch}
\APACinsertmetastar {%
buzzicotti2020synch}%
\begin{APACrefauthors}%
Buzzicotti, M.%
\BCBT {}\ \BBA {} Clark Di~Leoni, P.%
\end{APACrefauthors}%
\unskip\
\newblock
\APACrefYearMonthDay{2020}{}{}.
\newblock
{\BBOQ}\APACrefatitle {Synchronizing subgrid scale models of turbulence to
  data} {Synchronizing subgrid scale models of turbulence to data}.{\BBCQ}
\newblock
\APACjournalVolNumPages{Physics of Fluids}{32}{12}{125116}.
\PrintBackRefs{\CurrentBib}

\bibitem [\protect \citeauthoryear {%
Buzzicotti%
\ \protect \BOthers {.}}{%
Buzzicotti%
\ \protect \BOthers {.}}{%
{\protect \APACyear {2018}}%
}]{%
buzzicotti2018effect}
\APACinsertmetastar {%
buzzicotti2018effect}%
\begin{APACrefauthors}%
Buzzicotti, M.%
, Linkmann, M.%
, Aluie, H.%
, Biferale, L.%
, Brasseur, J.%
\BCBL {}\ \BBA {} Meneveau, C.%
\end{APACrefauthors}%
\unskip\
\newblock
\APACrefYearMonthDay{2018}{}{}.
\newblock
{\BBOQ}\APACrefatitle {Effect of filter type on the statistics of energy
  transfer between resolved and subfilter scales from a-priori analysis of
  direct numerical simulations of isotropic turbulence} {Effect of filter type
  on the statistics of energy transfer between resolved and subfilter scales
  from a-priori analysis of direct numerical simulations of isotropic
  turbulence}.{\BBCQ}
\newblock
\APACjournalVolNumPages{Journal of Turbulence}{19}{2}{167--197}.
\PrintBackRefs{\CurrentBib}

\bibitem [\protect \citeauthoryear {%
Buzzicotti%
\ \BBA {} Tauzin%
}{%
Buzzicotti%
\ \BBA {} Tauzin%
}{%
{\protect \APACyear {2021}}%
}]{%
buzzicotti2021inertial}
\APACinsertmetastar {%
buzzicotti2021inertial}%
\begin{APACrefauthors}%
Buzzicotti, M.%
\BCBT {}\ \BBA {} Tauzin, G.%
\end{APACrefauthors}%
\unskip\
\newblock
\APACrefYearMonthDay{2021}{}{}.
\newblock
{\BBOQ}\APACrefatitle {Inertial range statistics of the entropic lattice
  Boltzmann method in three-dimensional turbulence} {Inertial range statistics
  of the entropic lattice boltzmann method in three-dimensional
  turbulence}.{\BBCQ}
\newblock
\APACjournalVolNumPages{Physical Review E}{104}{1}{015302}.
\PrintBackRefs{\CurrentBib}

\bibitem [\protect \citeauthoryear {%
Callies%
\ \BBA {} Wu%
}{%
Callies%
\ \BBA {} Wu%
}{%
{\protect \APACyear {2019}}%
}]{%
CalliesWu2019}
\APACinsertmetastar {%
CalliesWu2019}%
\begin{APACrefauthors}%
Callies, J.%
\BCBT {}\ \BBA {} Wu, W.%
\end{APACrefauthors}%
\unskip\
\newblock
\APACrefYearMonthDay{2019}{{\APACmonth{09}}}{}.
\newblock
{\BBOQ}\APACrefatitle {{Some Expectations for Submesoscale Sea Surface Height
  Variance Spectra}} {{Some Expectations for Submesoscale Sea Surface Height
  Variance Spectra}}.{\BBCQ}
\newblock
\APACjournalVolNumPages{Journal of Physical Oceanography}{49}{9}{2271--2289}.
\PrintBackRefs{\CurrentBib}

\bibitem [\protect \citeauthoryear {%
Chen%
, Gille%
, McClean%
, Flierl%
\BCBL {}\ \BBA {} Griesel%
}{%
Chen%
\ \protect \BOthers {.}}{%
{\protect \APACyear {2015}}%
}]{%
Chenetal2015}
\APACinsertmetastar {%
Chenetal2015}%
\begin{APACrefauthors}%
Chen, R.%
, Gille, S\BPBI T.%
, McClean, J\BPBI L.%
, Flierl, G\BPBI R.%
\BCBL {}\ \BBA {} Griesel, A.%
\end{APACrefauthors}%
\unskip\
\newblock
\APACrefYearMonthDay{2015}{{\APACmonth{07}}}{}.
\newblock
{\BBOQ}\APACrefatitle {{A Multiwavenumber Theory for Eddy Diffusivities and Its
  Application to the Southeast Pacific (DIMES) Region}} {{A Multiwavenumber
  Theory for Eddy Diffusivities and Its Application to the Southeast Pacific
  (DIMES) Region}}.{\BBCQ}
\newblock
\APACjournalVolNumPages{Journal of Physical Oceanography}{45}{7}{1877--1896}.
\PrintBackRefs{\CurrentBib}

\bibitem [\protect \citeauthoryear {%
Di~Leoni%
, Alexakis%
, Biferale%
\BCBL {}\ \BBA {} Buzzicotti%
}{%
Di~Leoni%
\ \protect \BOthers {.}}{%
{\protect \APACyear {2020}}%
}]{%
di2020phase}
\APACinsertmetastar {%
di2020phase}%
\begin{APACrefauthors}%
Di~Leoni, P\BPBI C.%
, Alexakis, A.%
, Biferale, L.%
\BCBL {}\ \BBA {} Buzzicotti, M.%
\end{APACrefauthors}%
\unskip\
\newblock
\APACrefYearMonthDay{2020}{}{}.
\newblock
{\BBOQ}\APACrefatitle {Phase transitions and flux-loop metastable states in
  rotating turbulence} {Phase transitions and flux-loop metastable states in
  rotating turbulence}.{\BBCQ}
\newblock
\APACjournalVolNumPages{Physical Review Fluids}{5}{10}{104603}.
\PrintBackRefs{\CurrentBib}

\bibitem [\protect \citeauthoryear {%
Di~Lorenzo%
\ \protect \BOthers {.}}{%
Di~Lorenzo%
\ \protect \BOthers {.}}{%
{\protect \APACyear {2008}}%
}]{%
DiLorenzo2008}
\APACinsertmetastar {%
DiLorenzo2008}%
\begin{APACrefauthors}%
Di~Lorenzo, E.%
, Schneider, N.%
, Cobb, K\BPBI M.%
, Franks, P\BPBI J\BPBI S.%
, Chhak, K.%
, Miller, A\BPBI J.%
\BDBL {}Rivi{\`e}re, P.%
\end{APACrefauthors}%
\unskip\
\newblock
\APACrefYearMonthDay{2008}{{\APACmonth{04}}}{}.
\newblock
{\BBOQ}\APACrefatitle {{North Pacific Gyre Oscillation links ocean climate and
  ecosystem change}} {{North Pacific Gyre Oscillation links ocean climate and
  ecosystem change}}.{\BBCQ}
\newblock
\APACjournalVolNumPages{Geophysical Research Letters}{35}{8}{L08607}.
\PrintBackRefs{\CurrentBib}

\bibitem [\protect \citeauthoryear {%
Eyink%
\ \BBA {} Aluie%
}{%
Eyink%
\ \BBA {} Aluie%
}{%
{\protect \APACyear {2009}}%
}]{%
EyinkAluie09}
\APACinsertmetastar {%
EyinkAluie09}%
\begin{APACrefauthors}%
Eyink, G.%
\BCBT {}\ \BBA {} Aluie, H.%
\end{APACrefauthors}%
\unskip\
\newblock
\APACrefYearMonthDay{2009}{{\APACmonth{11}}}{}.
\newblock
{\BBOQ}\APACrefatitle {{Localness of energy cascade in hydrodynamic turbulence.
  I. Smooth coarse graining}} {{Localness of energy cascade in hydrodynamic
  turbulence. I. Smooth coarse graining}}.{\BBCQ}
\newblock
\APACjournalVolNumPages{Phys. Fluids}{21}{11}{115107}.
\PrintBackRefs{\CurrentBib}

\bibitem [\protect \citeauthoryear {%
{Eyink}%
}{%
{Eyink}%
}{%
{\protect \APACyear {1995}}%
}]{%
Eyink95}
\APACinsertmetastar {%
Eyink95}%
\begin{APACrefauthors}%
{Eyink}, G\BPBI L.%
\end{APACrefauthors}%
\unskip\
\newblock
\APACrefYearMonthDay{1995}{}{}.
\newblock
{\BBOQ}\APACrefatitle {{Local energy flux and the refined similarity
  hypothesis}} {{Local energy flux and the refined similarity
  hypothesis}}.{\BBCQ}
\newblock
\APACjournalVolNumPages{J. Stat. Phys.}{78}{}{335-351}.
\newblock
\begin{APACrefDOI} \doi{10.1007/BF02183352} \end{APACrefDOI}
\PrintBackRefs{\CurrentBib}

\bibitem [\protect \citeauthoryear {%
{Eyink}%
}{%
{Eyink}%
}{%
{\protect \APACyear {2005}}%
}]{%
Eyink05}
\APACinsertmetastar {%
Eyink05}%
\begin{APACrefauthors}%
{Eyink}, G\BPBI L.%
\end{APACrefauthors}%
\unskip\
\newblock
\APACrefYearMonthDay{2005}{}{}.
\newblock
{\BBOQ}\APACrefatitle {{Locality of turbulent cascades}} {{Locality of
  turbulent cascades}}.{\BBCQ}
\newblock
\APACjournalVolNumPages{Physica D}{207}{}{91-116}.
\newblock
\begin{APACrefDOI} \doi{10.1016/j.physd.2005.05.018} \end{APACrefDOI}
\PrintBackRefs{\CurrentBib}

\bibitem [\protect \citeauthoryear {%
Ferrari%
\ \BBA {} Wunsch%
}{%
Ferrari%
\ \BBA {} Wunsch%
}{%
{\protect \APACyear {2009}}%
}]{%
FerrariWunsch09}
\APACinsertmetastar {%
FerrariWunsch09}%
\begin{APACrefauthors}%
Ferrari, R.%
\BCBT {}\ \BBA {} Wunsch, C.%
\end{APACrefauthors}%
\unskip\
\newblock
\APACrefYearMonthDay{2009}{{\APACmonth{01}}}{}.
\newblock
{\BBOQ}\APACrefatitle {{Ocean Circulation Kinetic Energy: Reservoirs, Sources,
  and Sinks}} {{Ocean Circulation Kinetic Energy: Reservoirs, Sources, and
  Sinks}}.{\BBCQ}
\newblock
\APACjournalVolNumPages{Annual Review of Fluid Mechanics}{41}{1}{253--282}.
\PrintBackRefs{\CurrentBib}

\bibitem [\protect \citeauthoryear {%
Fox-Kemper%
\ \protect \BOthers {.}}{%
Fox-Kemper%
\ \protect \BOthers {.}}{%
{\protect \APACyear {2011}}%
}]{%
fox2011parameterization}
\APACinsertmetastar {%
fox2011parameterization}%
\begin{APACrefauthors}%
Fox-Kemper, B.%
, Danabasoglu, G.%
, Ferrari, R.%
, Griffies, S.%
, Hallberg, R.%
, Holland, M.%
\BDBL {}Samuels, B.%
\end{APACrefauthors}%
\unskip\
\newblock
\APACrefYearMonthDay{2011}{}{}.
\newblock
{\BBOQ}\APACrefatitle {Parameterization of mixed layer eddies. III:
  Implementation and impact in global ocean climate simulations}
  {Parameterization of mixed layer eddies. iii: Implementation and impact in
  global ocean climate simulations}.{\BBCQ}
\newblock
\APACjournalVolNumPages{Ocean Modelling}{39}{1-2}{61--78}.
\PrintBackRefs{\CurrentBib}

\bibitem [\protect \citeauthoryear {%
Fu%
\ \BBA {} Smith%
}{%
Fu%
\ \BBA {} Smith%
}{%
{\protect \APACyear {1996}}%
}]{%
FuSmith1996}
\APACinsertmetastar {%
FuSmith1996}%
\begin{APACrefauthors}%
Fu, L\BHBI L.%
\BCBT {}\ \BBA {} Smith, R\BPBI D.%
\end{APACrefauthors}%
\unskip\
\newblock
\APACrefYearMonthDay{1996}{{\APACmonth{11}}}{}.
\newblock
{\BBOQ}\APACrefatitle {{Global Ocean Circulation from Satellite Altimetry and
  High-Resolution Computer Simulation.}} {{Global Ocean Circulation from
  Satellite Altimetry and High-Resolution Computer Simulation.}}{\BBCQ}
\newblock
\APACjournalVolNumPages{Bulletin of the American Meteorological
  Society}{77}{1}{2625--2636}.
\PrintBackRefs{\CurrentBib}

\bibitem [\protect \citeauthoryear {%
{Germano}%
}{%
{Germano}%
}{%
{\protect \APACyear {1992}}%
}]{%
Germano92}
\APACinsertmetastar {%
Germano92}%
\begin{APACrefauthors}%
{Germano}, M.%
\end{APACrefauthors}%
\unskip\
\newblock
\APACrefYearMonthDay{1992}{}{}.
\newblock
{\BBOQ}\APACrefatitle {{Turbulence - The filtering approach}} {{Turbulence -
  The filtering approach}}.{\BBCQ}
\newblock
\APACjournalVolNumPages{J. Fluid Mech.}{238}{}{325-336}.
\newblock
\begin{APACrefDOI} \doi{10.1017/S0022112092001733} \end{APACrefDOI}
\PrintBackRefs{\CurrentBib}

\bibitem [\protect \citeauthoryear {%
Germano%
}{%
Germano%
}{%
{\protect \APACyear {1992}}%
}]{%
germano1992turbulence}
\APACinsertmetastar {%
germano1992turbulence}%
\begin{APACrefauthors}%
Germano, M.%
\end{APACrefauthors}%
\unskip\
\newblock
\APACrefYearMonthDay{1992}{}{}.
\newblock
{\BBOQ}\APACrefatitle {Turbulence: the filtering approach} {Turbulence: the
  filtering approach}.{\BBCQ}
\newblock
\APACjournalVolNumPages{Journal of Fluid Mechanics}{238}{}{325--336}.
\PrintBackRefs{\CurrentBib}

\bibitem [\protect \citeauthoryear {%
Griffies%
\ \protect \BOthers {.}}{%
Griffies%
\ \protect \BOthers {.}}{%
{\protect \APACyear {2015}}%
}]{%
Griffiesetal15}
\APACinsertmetastar {%
Griffiesetal15}%
\begin{APACrefauthors}%
Griffies, S\BPBI M.%
, Winton, M.%
, Anderson, W\BPBI G.%
, Benson, R.%
, Delworth, T\BPBI L.%
, Dufour, C\BPBI O.%
\BDBL {}Zhang, R.%
\end{APACrefauthors}%
\unskip\
\newblock
\APACrefYearMonthDay{2015}{{\APACmonth{02}}}{}.
\newblock
{\BBOQ}\APACrefatitle {{Impacts on Ocean Heat from Transient Mesoscale Eddies
  in a Hierarchy of Climate Models}} {{Impacts on Ocean Heat from Transient
  Mesoscale Eddies in a Hierarchy of Climate Models}}.{\BBCQ}
\newblock
\APACjournalVolNumPages{Journal of Climate}{28}{3}{952--977}.
\PrintBackRefs{\CurrentBib}

\bibitem [\protect \citeauthoryear {%
Grooms%
\ \protect \BOthers {.}}{%
Grooms%
\ \protect \BOthers {.}}{%
{\protect \APACyear {2021}}%
}]{%
grooms_etal2021}
\APACinsertmetastar {%
grooms_etal2021}%
\begin{APACrefauthors}%
Grooms, I.%
, Loose, N.%
, Abernathey, R.%
, Steinberg, J.%
, Bachman, S.%
, Marques, G.%
\BDBL {}Yankovsky, E.%
\end{APACrefauthors}%
\unskip\
\newblock
\APACrefYearMonthDay{2021}{}{}.
\newblock
{\BBOQ}\APACrefatitle {Diffusion-based smoothers for spatial filtering of
  gridded geophysical data} {Diffusion-based smoothers for spatial filtering of
  gridded geophysical data}.{\BBCQ}
\newblock
\APACjournalVolNumPages{Journal of Advances in Modeling the Earth
  System}{submitted}{}{}.
\PrintBackRefs{\CurrentBib}

\bibitem [\protect \citeauthoryear {%
Haigh%
, Sun%
, McWilliams%
\BCBL {}\ \BBA {} Berloff%
}{%
Haigh%
\ \protect \BOthers {.}}{%
{\protect \APACyear {2021}}%
}]{%
haigh2021eddy}
\APACinsertmetastar {%
haigh2021eddy}%
\begin{APACrefauthors}%
Haigh, M.%
, Sun, L.%
, McWilliams, J\BPBI C.%
\BCBL {}\ \BBA {} Berloff, P.%
\end{APACrefauthors}%
\unskip\
\newblock
\APACrefYearMonthDay{2021}{}{}.
\newblock
{\BBOQ}\APACrefatitle {On eddy transport in the ocean. Part I: The diffusion
  tensor} {On eddy transport in the ocean. part i: The diffusion
  tensor}.{\BBCQ}
\newblock
\APACjournalVolNumPages{Ocean Modelling}{164}{}{101831}.
\PrintBackRefs{\CurrentBib}

\bibitem [\protect \citeauthoryear {%
Haigh%
, Sun%
, Shevchenko%
\BCBL {}\ \BBA {} Berloff%
}{%
Haigh%
\ \protect \BOthers {.}}{%
{\protect \APACyear {2020}}%
}]{%
haigh2020tracer}
\APACinsertmetastar {%
haigh2020tracer}%
\begin{APACrefauthors}%
Haigh, M.%
, Sun, L.%
, Shevchenko, I.%
\BCBL {}\ \BBA {} Berloff, P.%
\end{APACrefauthors}%
\unskip\
\newblock
\APACrefYearMonthDay{2020}{}{}.
\newblock
{\BBOQ}\APACrefatitle {Tracer-based estimates of eddy-induced diffusivities}
  {Tracer-based estimates of eddy-induced diffusivities}.{\BBCQ}
\newblock
\APACjournalVolNumPages{Deep Sea Research Part I: Oceanographic Research
  Papers}{160}{}{103264}.
\PrintBackRefs{\CurrentBib}

\bibitem [\protect \citeauthoryear {%
Hewitt%
\ \protect \BOthers {.}}{%
Hewitt%
\ \protect \BOthers {.}}{%
{\protect \APACyear {2011}}%
}]{%
gmd-4-223-2011}
\APACinsertmetastar {%
gmd-4-223-2011}%
\begin{APACrefauthors}%
Hewitt, H\BPBI T.%
, Copsey, D.%
, Culverwell, I\BPBI D.%
, Harris, C\BPBI M.%
, Hill, R\BPBI S\BPBI R.%
, Keen, A\BPBI B.%
\BDBL {}Hunke, E\BPBI C.%
\end{APACrefauthors}%
\unskip\
\newblock
\APACrefYearMonthDay{2011}{}{}.
\newblock
{\BBOQ}\APACrefatitle {Design and implementation of the infrastructure of
  HadGEM3: the next-generation Met Office climate modelling system} {Design and
  implementation of the infrastructure of hadgem3: the next-generation met
  office climate modelling system}.{\BBCQ}
\newblock
\APACjournalVolNumPages{Geoscientific Model Development}{4}{2}{223--253}.
\newblock
\begin{APACrefDOI} \doi{10.5194/gmd-4-223-2011} \end{APACrefDOI}
\PrintBackRefs{\CurrentBib}

\bibitem [\protect \citeauthoryear {%
Jansen%
, Adcroft%
, Khani%
\BCBL {}\ \BBA {} Kong%
}{%
Jansen%
\ \protect \BOthers {.}}{%
{\protect \APACyear {2019}}%
}]{%
jansen2019toward}
\APACinsertmetastar {%
jansen2019toward}%
\begin{APACrefauthors}%
Jansen, M\BPBI F.%
, Adcroft, A.%
, Khani, S.%
\BCBL {}\ \BBA {} Kong, H.%
\end{APACrefauthors}%
\unskip\
\newblock
\APACrefYearMonthDay{2019}{}{}.
\newblock
{\BBOQ}\APACrefatitle {Toward an energetically consistent, resolution aware
  parameterization of ocean mesoscale eddies} {Toward an energetically
  consistent, resolution aware parameterization of ocean mesoscale
  eddies}.{\BBCQ}
\newblock
\APACjournalVolNumPages{Journal of Advances in Modeling Earth
  Systems}{11}{8}{2844--2860}.
\PrintBackRefs{\CurrentBib}

\bibitem [\protect \citeauthoryear {%
Kac%
\ \BBA {} Siegert%
}{%
Kac%
\ \BBA {} Siegert%
}{%
{\protect \APACyear {1947}}%
}]{%
KacSiegert1947}
\APACinsertmetastar {%
KacSiegert1947}%
\begin{APACrefauthors}%
Kac, M.%
\BCBT {}\ \BBA {} Siegert, A.%
\end{APACrefauthors}%
\unskip\
\newblock
\APACrefYearMonthDay{1947}{}{}.
\newblock
{\BBOQ}\APACrefatitle {An explicit representation of a stationary Gaussian
  process} {An explicit representation of a stationary gaussian
  process}.{\BBCQ}
\newblock
\APACjournalVolNumPages{The Annals of Mathematical
  Statistics}{18}{3}{438--442}.
\PrintBackRefs{\CurrentBib}

\bibitem [\protect \citeauthoryear {%
Karhunen%
}{%
Karhunen%
}{%
{\protect \APACyear {1947}}%
}]{%
Karhunen1947}
\APACinsertmetastar {%
Karhunen1947}%
\begin{APACrefauthors}%
Karhunen, K.%
\end{APACrefauthors}%
\unskip\
\newblock
\APACrefYearMonthDay{1947}{}{}.
\newblock
{\BBOQ}\APACrefatitle {Under lineare methoden in der wahr
  scheinlichkeitsrechnung} {Under lineare methoden in der wahr
  scheinlichkeitsrechnung}.{\BBCQ}
\newblock
\APACjournalVolNumPages{Annales Academiae Scientiarun Fennicae Series A1:
  Mathematia Physica}{47}{}{}.
\PrintBackRefs{\CurrentBib}

\bibitem [\protect \citeauthoryear {%
Kawabe%
}{%
Kawabe%
}{%
{\protect \APACyear {1995}}%
}]{%
kawabe1995variations}
\APACinsertmetastar {%
kawabe1995variations}%
\begin{APACrefauthors}%
Kawabe, M.%
\end{APACrefauthors}%
\unskip\
\newblock
\APACrefYearMonthDay{1995}{}{}.
\newblock
{\BBOQ}\APACrefatitle {Variations of current path, velocity, and volume
  transport of the Kuroshio in relation with the large meander} {Variations of
  current path, velocity, and volume transport of the kuroshio in relation with
  the large meander}.{\BBCQ}
\newblock
\APACjournalVolNumPages{Journal of physical oceanography}{25}{12}{3103--3117}.
\PrintBackRefs{\CurrentBib}

\bibitem [\protect \citeauthoryear {%
Kelley%
\ \BBA {} Ouellette%
}{%
Kelley%
\ \BBA {} Ouellette%
}{%
{\protect \APACyear {2011}}%
}]{%
KelleyOuellette11}
\APACinsertmetastar {%
KelleyOuellette11}%
\begin{APACrefauthors}%
Kelley, D\BPBI H.%
\BCBT {}\ \BBA {} Ouellette, N\BPBI T.%
\end{APACrefauthors}%
\unskip\
\newblock
\APACrefYearMonthDay{2011}{{\APACmonth{11}}}{}.
\newblock
{\BBOQ}\APACrefatitle {{Spatiotemporal persistence of spectral fluxes in
  two-dimensional weak turbulence}} {{Spatiotemporal persistence of spectral
  fluxes in two-dimensional weak turbulence}}.{\BBCQ}
\newblock
\APACjournalVolNumPages{Physics of Fluids}{23}{1}{5101}.
\PrintBackRefs{\CurrentBib}

\bibitem [\protect \citeauthoryear {%
Kessler%
}{%
Kessler%
}{%
{\protect \APACyear {1990}}%
}]{%
kessler1990observations}
\APACinsertmetastar {%
kessler1990observations}%
\begin{APACrefauthors}%
Kessler, W\BPBI S.%
\end{APACrefauthors}%
\unskip\
\newblock
\APACrefYearMonthDay{1990}{}{}.
\newblock
{\BBOQ}\APACrefatitle {Observations of long Rossby waves in the northern
  tropical Pacific} {Observations of long rossby waves in the northern tropical
  pacific}.{\BBCQ}
\newblock
\APACjournalVolNumPages{Journal of Geophysical Research:
  Oceans}{95}{C4}{5183--5217}.
\PrintBackRefs{\CurrentBib}

\bibitem [\protect \citeauthoryear {%
Khani%
\ \BBA {} Dawson%
}{%
Khani%
\ \BBA {} Dawson%
}{%
{\protect \APACyear {2023}}%
}]{%
khani2023gradient}
\APACinsertmetastar {%
khani2023gradient}%
\begin{APACrefauthors}%
Khani, S.%
\BCBT {}\ \BBA {} Dawson, C\BPBI N.%
\end{APACrefauthors}%
\unskip\
\newblock
\APACrefYearMonthDay{2023}{}{}.
\newblock
{\BBOQ}\APACrefatitle {A gradient based subgrid-scale parameterization for
  ocean mesoscale eddies} {A gradient based subgrid-scale parameterization for
  ocean mesoscale eddies}.{\BBCQ}
\newblock
\APACjournalVolNumPages{Journal of Advances in Modeling Earth
  Systems}{15}{2}{e2022MS003356}.
\PrintBackRefs{\CurrentBib}

\bibitem [\protect \citeauthoryear {%
Khani%
, Jansen%
\BCBL {}\ \BBA {} Adcroft%
}{%
Khani%
\ \protect \BOthers {.}}{%
{\protect \APACyear {2019}}%
}]{%
khani2019diagnosing}
\APACinsertmetastar {%
khani2019diagnosing}%
\begin{APACrefauthors}%
Khani, S.%
, Jansen, M\BPBI F.%
\BCBL {}\ \BBA {} Adcroft, A.%
\end{APACrefauthors}%
\unskip\
\newblock
\APACrefYearMonthDay{2019}{}{}.
\newblock
{\BBOQ}\APACrefatitle {Diagnosing subgrid mesoscale eddy fluxes with and
  without topography} {Diagnosing subgrid mesoscale eddy fluxes with and
  without topography}.{\BBCQ}
\newblock
\APACjournalVolNumPages{Journal of Advances in Modeling Earth
  Systems}{11}{12}{3995--4015}.
\PrintBackRefs{\CurrentBib}

\bibitem [\protect \citeauthoryear {%
Khatri%
, Sukhatme%
, Kumar%
\BCBL {}\ \BBA {} Verma%
}{%
Khatri%
\ \protect \BOthers {.}}{%
{\protect \APACyear {2018}}%
}]{%
khatri2018surface}
\APACinsertmetastar {%
khatri2018surface}%
\begin{APACrefauthors}%
Khatri, H.%
, Sukhatme, J.%
, Kumar, A.%
\BCBL {}\ \BBA {} Verma, M\BPBI K.%
\end{APACrefauthors}%
\unskip\
\newblock
\APACrefYearMonthDay{2018}{}{}.
\newblock
{\BBOQ}\APACrefatitle {Surface ocean enstrophy, kinetic energy fluxes, and
  spectra from satellite altimetry} {Surface ocean enstrophy, kinetic energy
  fluxes, and spectra from satellite altimetry}.{\BBCQ}
\newblock
\APACjournalVolNumPages{Journal of Geophysical Research:
  Oceans}{123}{5}{3875--3892}.
\newblock
\begin{APACrefDOI} \doi{10.1029/2017JC013516} \end{APACrefDOI}
\PrintBackRefs{\CurrentBib}

\bibitem [\protect \citeauthoryear {%
Lea%
\ \protect \BOthers {.}}{%
Lea%
\ \protect \BOthers {.}}{%
{\protect \APACyear {2015}}%
}]{%
lea2015assessing}
\APACinsertmetastar {%
lea2015assessing}%
\begin{APACrefauthors}%
Lea, D.%
, Mirouze, I.%
, Martin, M.%
, King, R.%
, Hines, A.%
, Walters, D.%
\BCBL {}\ \BBA {} Thurlow, M.%
\end{APACrefauthors}%
\unskip\
\newblock
\APACrefYearMonthDay{2015}{}{}.
\newblock
{\BBOQ}\APACrefatitle {Assessing a new coupled data assimilation system based
  on the Met Office coupled atmosphere--land--ocean--sea ice model} {Assessing
  a new coupled data assimilation system based on the met office coupled
  atmosphere--land--ocean--sea ice model}.{\BBCQ}
\newblock
\APACjournalVolNumPages{Monthly Weather Review}{143}{11}{4678--4694}.
\PrintBackRefs{\CurrentBib}

\bibitem [\protect \citeauthoryear {%
{Leonard}%
}{%
{Leonard}%
}{%
{\protect \APACyear {1974}}%
}]{%
Leonard74}
\APACinsertmetastar {%
Leonard74}%
\begin{APACrefauthors}%
{Leonard}, A.%
\end{APACrefauthors}%
\unskip\
\newblock
\APACrefYearMonthDay{1974}{}{}.
\newblock
{\BBOQ}\APACrefatitle {{Energy Cascade in Large-Eddy Simulations of Turbulent
  Fluid Flows}} {{Energy Cascade in Large-Eddy Simulations of Turbulent Fluid
  Flows}}.{\BBCQ}
\newblock
\APACjournalVolNumPages{Adv. Geophys.}{18}{}{A237}.
\PrintBackRefs{\CurrentBib}

\bibitem [\protect \citeauthoryear {%
Lieb%
\ \BBA {} Loss%
}{%
Lieb%
\ \BBA {} Loss%
}{%
{\protect \APACyear {2001}}%
}]{%
LiebLoss2001}
\APACinsertmetastar {%
LiebLoss2001}%
\begin{APACrefauthors}%
Lieb, E\BPBI H.%
\BCBT {}\ \BBA {} Loss, M.%
\end{APACrefauthors}%
\unskip\
\newblock
\APACrefYearMonthDay{2001}{}{}.
\newblock
{\BBOQ}\APACrefatitle {Analysis 2nd eq.} {Analysis 2nd eq.}{\BBCQ}
\newblock
\APACjournalVolNumPages{American Mathematical Society, Providence,
  RI}{14}{}{348}.
\PrintBackRefs{\CurrentBib}

\bibitem [\protect \citeauthoryear {%
Linkmann%
, Buzzicotti%
\BCBL {}\ \BBA {} Biferale%
}{%
Linkmann%
\ \protect \BOthers {.}}{%
{\protect \APACyear {2018}}%
}]{%
linkmann2018multi}
\APACinsertmetastar {%
linkmann2018multi}%
\begin{APACrefauthors}%
Linkmann, M.%
, Buzzicotti, M.%
\BCBL {}\ \BBA {} Biferale, L.%
\end{APACrefauthors}%
\unskip\
\newblock
\APACrefYearMonthDay{2018}{}{}.
\newblock
{\BBOQ}\APACrefatitle {Multi-scale properties of large eddy simulations:
  correlations between resolved-scale velocity-field increments and
  subgrid-scale quantities} {Multi-scale properties of large eddy simulations:
  correlations between resolved-scale velocity-field increments and
  subgrid-scale quantities}.{\BBCQ}
\newblock
\APACjournalVolNumPages{Journal of Turbulence}{19}{6}{493--527}.
\PrintBackRefs{\CurrentBib}

\bibitem [\protect \citeauthoryear {%
Loeve%
}{%
Loeve%
}{%
{\protect \APACyear {1948}}%
}]{%
Loeve1948}
\APACinsertmetastar {%
Loeve1948}%
\begin{APACrefauthors}%
Loeve, M.%
\end{APACrefauthors}%
\unskip\
\newblock
\APACrefYearMonthDay{1948}{}{}.
\newblock
{\BBOQ}\APACrefatitle {Functions aleatoires du second ordre} {Functions
  aleatoires du second ordre}.{\BBCQ}
\newblock
\BIn{} P.~Levy\ (\BED), \APACrefbtitle {Processus stochastique et mouvement
  Brownien} {Processus stochastique et mouvement brownien}\ (\BPGS\ 366--420).
\newblock
\APACaddressPublisher{Paris}{Gauthier-Villars}.
\PrintBackRefs{\CurrentBib}

\bibitem [\protect \citeauthoryear {%
Loose%
, Bachman%
, Grooms%
\BCBL {}\ \BBA {} Jansen%
}{%
Loose%
\ \protect \BOthers {.}}{%
{\protect \APACyear {2023}}%
}]{%
loose2023diagnosing}
\APACinsertmetastar {%
loose2023diagnosing}%
\begin{APACrefauthors}%
Loose, N.%
, Bachman, S.%
, Grooms, I.%
\BCBL {}\ \BBA {} Jansen, M.%
\end{APACrefauthors}%
\unskip\
\newblock
\APACrefYearMonthDay{2023}{}{}.
\newblock
{\BBOQ}\APACrefatitle {Diagnosing scale-dependent energy cycles in a
  high-resolution isopycnal ocean model} {Diagnosing scale-dependent energy
  cycles in a high-resolution isopycnal ocean model}.{\BBCQ}
\newblock
\APACjournalVolNumPages{Journal of Physical Oceanography}{53}{1}{157--176}.
\PrintBackRefs{\CurrentBib}

\bibitem [\protect \citeauthoryear {%
Lorenz%
}{%
Lorenz%
}{%
{\protect \APACyear {1956}}%
}]{%
Lorenz1956}
\APACinsertmetastar {%
Lorenz1956}%
\begin{APACrefauthors}%
Lorenz, E\BPBI N.%
\end{APACrefauthors}%
\unskip\
\newblock
\APACrefYear{1956}.
\newblock
\APACrefbtitle {Empirical orthogonal functions and statistical weather
  prediction} {Empirical orthogonal functions and statistical weather
  prediction}\ (\BVOL~1).
\newblock
\APACaddressPublisher{}{Massachusetts Institute of Technology, Department of
  Meteorology Cambridge}.
\PrintBackRefs{\CurrentBib}

\bibitem [\protect \citeauthoryear {%
Meneveau%
}{%
Meneveau%
}{%
{\protect \APACyear {1994}}%
}]{%
Meneveau1994}
\APACinsertmetastar {%
Meneveau1994}%
\begin{APACrefauthors}%
Meneveau, C.%
\end{APACrefauthors}%
\unskip\
\newblock
\APACrefYearMonthDay{1994}{{\APACmonth{02}}}{}.
\newblock
{\BBOQ}\APACrefatitle {{Statistics of Turbulence Subgrid-Scale Stresses -
  Necessary Conditions and Experimental Tests}} {{Statistics of Turbulence
  Subgrid-Scale Stresses - Necessary Conditions and Experimental
  Tests}}.{\BBCQ}
\newblock
\APACjournalVolNumPages{Physics of Fluids}{6}{2}{815--833}.
\PrintBackRefs{\CurrentBib}

\bibitem [\protect \citeauthoryear {%
Meneveau%
\ \BBA {} Katz%
}{%
Meneveau%
\ \BBA {} Katz%
}{%
{\protect \APACyear {2000}}%
}]{%
meneveau2000scale}
\APACinsertmetastar {%
meneveau2000scale}%
\begin{APACrefauthors}%
Meneveau, C.%
\BCBT {}\ \BBA {} Katz, J.%
\end{APACrefauthors}%
\unskip\
\newblock
\APACrefYearMonthDay{2000}{}{}.
\newblock
{\BBOQ}\APACrefatitle {Scale-invariance and turbulence models for large-eddy
  simulation} {Scale-invariance and turbulence models for large-eddy
  simulation}.{\BBCQ}
\newblock
\APACjournalVolNumPages{Annual Review of Fluid Mechanics}{32}{1}{1--32}.
\PrintBackRefs{\CurrentBib}

\bibitem [\protect \citeauthoryear {%
Muller%
\ \BBA {} Bony%
}{%
Muller%
\ \BBA {} Bony%
}{%
{\protect \APACyear {2015}}%
}]{%
muller2015favors}
\APACinsertmetastar {%
muller2015favors}%
\begin{APACrefauthors}%
Muller, C.%
\BCBT {}\ \BBA {} Bony, S.%
\end{APACrefauthors}%
\unskip\
\newblock
\APACrefYearMonthDay{2015}{}{}.
\newblock
{\BBOQ}\APACrefatitle {What favors convective aggregation and why?} {What
  favors convective aggregation and why?}{\BBCQ}
\newblock
\APACjournalVolNumPages{Geophysical Research Letters}{42}{13}{5626--5634}.
\PrintBackRefs{\CurrentBib}

\bibitem [\protect \citeauthoryear {%
O'Rourke%
, Arbic%
\BCBL {}\ \BBA {} Grif\/f\/ies%
}{%
O'Rourke%
\ \protect \BOthers {.}}{%
{\protect \APACyear {2018}}%
}]{%
ORourke_etal2018}
\APACinsertmetastar {%
ORourke_etal2018}%
\begin{APACrefauthors}%
O'Rourke, A\BPBI K.%
, Arbic, B.%
\BCBL {}\ \BBA {} Grif\/f\/ies, S.%
\end{APACrefauthors}%
\unskip\
\newblock
\APACrefYearMonthDay{2018}{}{}.
\newblock
{\BBOQ}\APACrefatitle {Frequency-domain analysis of atmospherically forced
  versus intrinsic ocean surface kinetic energy variability in {GFDL's CM2-O}
  model hierarchy} {Frequency-domain analysis of atmospherically forced versus
  intrinsic ocean surface kinetic energy variability in {GFDL's CM2-O} model
  hierarchy}.{\BBCQ}
\newblock
\APACjournalVolNumPages{Journal of Climate}{}{}{}.
\newblock
\begin{APACrefDOI} \doi{10.1175/JCLI-D-17-0024.1} \end{APACrefDOI}
\PrintBackRefs{\CurrentBib}

\bibitem [\protect \citeauthoryear {%
Pearson%
, Fox-Kemper%
, Bachman%
\BCBL {}\ \BBA {} Bryan%
}{%
Pearson%
\ \protect \BOthers {.}}{%
{\protect \APACyear {2017}}%
}]{%
Pearsonetal17}
\APACinsertmetastar {%
Pearsonetal17}%
\begin{APACrefauthors}%
Pearson, B.%
, Fox-Kemper, B.%
, Bachman, S.%
\BCBL {}\ \BBA {} Bryan, F.%
\end{APACrefauthors}%
\unskip\
\newblock
\APACrefYearMonthDay{2017}{{\APACmonth{07}}}{}.
\newblock
{\BBOQ}\APACrefatitle {{Evaluation of scale-aware subgrid mesoscale eddy models
  in a global eddy-rich model}} {{Evaluation of scale-aware subgrid mesoscale
  eddy models in a global eddy-rich model}}.{\BBCQ}
\newblock
\APACjournalVolNumPages{Ocean Modelling}{115}{}{42--58}.
\PrintBackRefs{\CurrentBib}

\bibitem [\protect \citeauthoryear {%
Piomelli%
, Cabot%
, Moin%
\BCBL {}\ \BBA {} Lee%
}{%
Piomelli%
\ \protect \BOthers {.}}{%
{\protect \APACyear {1991}}%
}]{%
Piomellietal91}
\APACinsertmetastar {%
Piomellietal91}%
\begin{APACrefauthors}%
Piomelli, U.%
, Cabot, W\BPBI H.%
, Moin, P.%
\BCBL {}\ \BBA {} Lee, S.%
\end{APACrefauthors}%
\unskip\
\newblock
\APACrefYearMonthDay{1991}{{\APACmonth{07}}}{}.
\newblock
{\BBOQ}\APACrefatitle {{Subgrid-scale backscatter in turbulent and transitional
  flows}} {{Subgrid-scale backscatter in turbulent and transitional
  flows}}.{\BBCQ}
\newblock
\APACjournalVolNumPages{Physics of Fluids A (ISSN 0899-8213)}{3}{}{1766--1771}.
\PrintBackRefs{\CurrentBib}

\bibitem [\protect \citeauthoryear {%
Pope%
}{%
Pope%
}{%
{\protect \APACyear {2001}}%
}]{%
pope2001turbulent}
\APACinsertmetastar {%
pope2001turbulent}%
\begin{APACrefauthors}%
Pope, S\BPBI B.%
\end{APACrefauthors}%
\unskip\
\newblock
\APACrefYearMonthDay{2001}{}{}.
\newblock
\APACrefbtitle {Turbulent flows.} {Turbulent flows.}
\newblock
\APACaddressPublisher{}{IOP Publishing}.
\PrintBackRefs{\CurrentBib}

\bibitem [\protect \citeauthoryear {%
Pujol%
\ \protect \BOthers {.}}{%
Pujol%
\ \protect \BOthers {.}}{%
{\protect \APACyear {2016}}%
}]{%
pujol2016}
\APACinsertmetastar {%
pujol2016}%
\begin{APACrefauthors}%
Pujol, M\BHBI I.%
, Faug{\`e}re, Y.%
, Taburet, G.%
, Dupuy, S.%
, Pelloquin, C.%
, Ablain, M.%
\BCBL {}\ \BBA {} Picot, N.%
\end{APACrefauthors}%
\unskip\
\newblock
\APACrefYearMonthDay{2016}{}{}.
\newblock
{\BBOQ}\APACrefatitle {DUACS DT2014: the new multi-mission altimeter data set
  reprocessed over 20 years} {Duacs dt2014: the new multi-mission altimeter
  data set reprocessed over 20 years}.{\BBCQ}
\newblock
\APACjournalVolNumPages{Ocean Sci}{12}{5}{1067--1090}.
\PrintBackRefs{\CurrentBib}

\bibitem [\protect \citeauthoryear {%
Qiu%
\ \protect \BOthers {.}}{%
Qiu%
\ \protect \BOthers {.}}{%
{\protect \APACyear {2018}}%
}]{%
Qiuetal2018jpo}
\APACinsertmetastar {%
Qiuetal2018jpo}%
\begin{APACrefauthors}%
Qiu, B.%
, Chen, S.%
, Klein, P.%
, Wang, J.%
, Torres, H.%
, Fu, L\BHBI L.%
\BCBL {}\ \BBA {} Menemenlis, D.%
\end{APACrefauthors}%
\unskip\
\newblock
\APACrefYearMonthDay{2018}{{\APACmonth{03}}}{}.
\newblock
{\BBOQ}\APACrefatitle {{Seasonality in Transition Scale from Balanced to
  Unbalanced Motions in the World Ocean}} {{Seasonality in Transition Scale
  from Balanced to Unbalanced Motions in the World Ocean}}.{\BBCQ}
\newblock
\APACjournalVolNumPages{Journal of Physical Oceanography}{48}{}{591--605}.
\PrintBackRefs{\CurrentBib}

\bibitem [\protect \citeauthoryear {%
Rai%
, Hecht%
, Maltrud%
\BCBL {}\ \BBA {} Aluie%
}{%
Rai%
\ \protect \BOthers {.}}{%
{\protect \APACyear {2021}}%
}]{%
Raietal2021}
\APACinsertmetastar {%
Raietal2021}%
\begin{APACrefauthors}%
Rai, S.%
, Hecht, M.%
, Maltrud, M.%
\BCBL {}\ \BBA {} Aluie, H.%
\end{APACrefauthors}%
\unskip\
\newblock
\APACrefYearMonthDay{2021}{}{}.
\newblock
{\BBOQ}\APACrefatitle {Scale of Oceanic Eddy-Killing by Wind from Global
  Satellite Observations} {Scale of oceanic eddy-killing by wind from global
  satellite observations}.{\BBCQ}
\newblock
\APACjournalVolNumPages{Science Advances}{}{}{}.
\newblock
\APACrefnote{under revision}
\PrintBackRefs{\CurrentBib}

\bibitem [\protect \citeauthoryear {%
Richman%
, Arbic%
, Shriver%
, Metzger%
\BCBL {}\ \BBA {} Wallcraft%
}{%
Richman%
\ \protect \BOthers {.}}{%
{\protect \APACyear {2012}}%
}]{%
richman2012inferring}
\APACinsertmetastar {%
richman2012inferring}%
\begin{APACrefauthors}%
Richman, J\BPBI G.%
, Arbic, B\BPBI K.%
, Shriver, J\BPBI F.%
, Metzger, E\BPBI J.%
\BCBL {}\ \BBA {} Wallcraft, A\BPBI J.%
\end{APACrefauthors}%
\unskip\
\newblock
\APACrefYearMonthDay{2012}{}{}.
\newblock
{\BBOQ}\APACrefatitle {Inferring dynamics from the wavenumber spectra of an
  eddying global ocean model with embedded tides} {Inferring dynamics from the
  wavenumber spectra of an eddying global ocean model with embedded
  tides}.{\BBCQ}
\newblock
\APACjournalVolNumPages{Journal of Geophysical Research: Oceans}{117}{C12}{}.
\PrintBackRefs{\CurrentBib}

\bibitem [\protect \citeauthoryear {%
Ringler%
\ \protect \BOthers {.}}{%
Ringler%
\ \protect \BOthers {.}}{%
{\protect \APACyear {2013}}%
}]{%
Ringleretal13}
\APACinsertmetastar {%
Ringleretal13}%
\begin{APACrefauthors}%
Ringler, T.%
, Petersen, M.%
, Higdon, R\BPBI L.%
, Jacobsen, D.%
, Jones, P\BPBI W.%
\BCBL {}\ \BBA {} Maltrud, M.%
\end{APACrefauthors}%
\unskip\
\newblock
\APACrefYearMonthDay{2013}{{\APACmonth{09}}}{}.
\newblock
{\BBOQ}\APACrefatitle {{A multi-resolution approach to global ocean modeling}}
  {{A multi-resolution approach to global ocean modeling}}.{\BBCQ}
\newblock
\APACjournalVolNumPages{Ocean Modelling}{69}{}{211--232}.
\PrintBackRefs{\CurrentBib}

\bibitem [\protect \citeauthoryear {%
Rivera%
, Aluie%
\BCBL {}\ \BBA {} Ecke%
}{%
Rivera%
\ \protect \BOthers {.}}{%
{\protect \APACyear {2014}}%
}]{%
Riveraetal14}
\APACinsertmetastar {%
Riveraetal14}%
\begin{APACrefauthors}%
Rivera, M\BPBI K.%
, Aluie, H.%
\BCBL {}\ \BBA {} Ecke, R\BPBI E.%
\end{APACrefauthors}%
\unskip\
\newblock
\APACrefYearMonthDay{2014}{{\APACmonth{05}}}{}.
\newblock
{\BBOQ}\APACrefatitle {{The direct enstrophy cascade of two-dimensional soap
  film flows}} {{The direct enstrophy cascade of two-dimensional soap film
  flows}}.{\BBCQ}
\newblock
\APACjournalVolNumPages{Physics of Fluids}{26}{5}{}.
\PrintBackRefs{\CurrentBib}

\bibitem [\protect \citeauthoryear {%
Rocha%
, Chereskin%
, Gille%
\BCBL {}\ \BBA {} Menemenlis%
}{%
Rocha%
\ \protect \BOthers {.}}{%
{\protect \APACyear {2016}}%
}]{%
Rochaetal2016}
\APACinsertmetastar {%
Rochaetal2016}%
\begin{APACrefauthors}%
Rocha, C\BPBI B.%
, Chereskin, T\BPBI K.%
, Gille, S\BPBI T.%
\BCBL {}\ \BBA {} Menemenlis, D.%
\end{APACrefauthors}%
\unskip\
\newblock
\APACrefYearMonthDay{2016}{{\APACmonth{02}}}{}.
\newblock
{\BBOQ}\APACrefatitle {{Mesoscale to Submesoscale Wavenumber Spectra in Drake
  Passage}} {{Mesoscale to Submesoscale Wavenumber Spectra in Drake
  Passage}}.{\BBCQ}
\newblock
\APACjournalVolNumPages{Journal of Physical Oceanography}{46}{}{601--620}.
\PrintBackRefs{\CurrentBib}

\bibitem [\protect \citeauthoryear {%
Ross%
, Li%
, Perezhogin%
, Fernandez-Granda%
\BCBL {}\ \BBA {} Zanna%
}{%
Ross%
\ \protect \BOthers {.}}{%
{\protect \APACyear {2023}}%
}]{%
ross2023benchmarking}
\APACinsertmetastar {%
ross2023benchmarking}%
\begin{APACrefauthors}%
Ross, A.%
, Li, Z.%
, Perezhogin, P.%
, Fernandez-Granda, C.%
\BCBL {}\ \BBA {} Zanna, L.%
\end{APACrefauthors}%
\unskip\
\newblock
\APACrefYearMonthDay{2023}{}{}.
\newblock
{\BBOQ}\APACrefatitle {Benchmarking of machine learning ocean subgrid
  parameterizations in an idealized model} {Benchmarking of machine learning
  ocean subgrid parameterizations in an idealized model}.{\BBCQ}
\newblock
\APACjournalVolNumPages{Journal of Advances in Modeling Earth
  Systems}{15}{1}{e2022MS003258}.
\PrintBackRefs{\CurrentBib}

\bibitem [\protect \citeauthoryear {%
Ryzhov%
, Kondrashov%
, Agarwal%
, McWilliams%
\BCBL {}\ \BBA {} Berloff%
}{%
Ryzhov%
\ \protect \BOthers {.}}{%
{\protect \APACyear {2020}}%
}]{%
ryzhov2020data}
\APACinsertmetastar {%
ryzhov2020data}%
\begin{APACrefauthors}%
Ryzhov, E.%
, Kondrashov, D.%
, Agarwal, N.%
, McWilliams, J.%
\BCBL {}\ \BBA {} Berloff, P.%
\end{APACrefauthors}%
\unskip\
\newblock
\APACrefYearMonthDay{2020}{}{}.
\newblock
{\BBOQ}\APACrefatitle {On data-driven induction of the low-frequency
  variability in a coarse-resolution ocean model} {On data-driven induction of
  the low-frequency variability in a coarse-resolution ocean model}.{\BBCQ}
\newblock
\APACjournalVolNumPages{Ocean Modelling}{153}{}{101664}.
\PrintBackRefs{\CurrentBib}

\bibitem [\protect \citeauthoryear {%
Sadek%
\ \BBA {} Aluie%
}{%
Sadek%
\ \BBA {} Aluie%
}{%
{\protect \APACyear {2018}}%
}]{%
sadek2018extracting}
\APACinsertmetastar {%
sadek2018extracting}%
\begin{APACrefauthors}%
Sadek, M.%
\BCBT {}\ \BBA {} Aluie, H.%
\end{APACrefauthors}%
\unskip\
\newblock
\APACrefYearMonthDay{2018}{}{}.
\newblock
{\BBOQ}\APACrefatitle {Extracting the spectrum of a flow by spatial filtering}
  {Extracting the spectrum of a flow by spatial filtering}.{\BBCQ}
\newblock
\APACjournalVolNumPages{Physical Review Fluids}{3}{12}{124610}.
\PrintBackRefs{\CurrentBib}

\bibitem [\protect \citeauthoryear {%
Satoh%
}{%
Satoh%
}{%
{\protect \APACyear {2004}}%
}]{%
satoh2004atmospheric}
\APACinsertmetastar {%
satoh2004atmospheric}%
\begin{APACrefauthors}%
Satoh, M.%
\end{APACrefauthors}%
\unskip\
\newblock
\APACrefYear{2004}.
\newblock
\APACrefbtitle {Atmospheric circulation dynamics and circulation models}
  {Atmospheric circulation dynamics and circulation models}.
\newblock
\APACaddressPublisher{}{Springer Science \& Business Media}.
\PrintBackRefs{\CurrentBib}

\bibitem [\protect \citeauthoryear {%
Savage%
\ \protect \BOthers {.}}{%
Savage%
\ \protect \BOthers {.}}{%
{\protect \APACyear {2017}}%
}]{%
Savageetal2017jgr}
\APACinsertmetastar {%
Savageetal2017jgr}%
\begin{APACrefauthors}%
Savage, A\BPBI C.%
, Arbic, B\BPBI K.%
, Alford, M\BPBI H.%
, Ansong, J\BPBI K.%
, Farrar, J\BPBI T.%
, Menemenlis, D.%
\BDBL {}Zamudio, L.%
\end{APACrefauthors}%
\unskip\
\newblock
\APACrefYearMonthDay{2017}{{\APACmonth{10}}}{}.
\newblock
{\BBOQ}\APACrefatitle {{Spectral decomposition of internal gravity wave sea
  surface height in global models}} {{Spectral decomposition of internal
  gravity wave sea surface height in global models}}.{\BBCQ}
\newblock
\APACjournalVolNumPages{Journal of Geophysical
  Research-Oceans}{122}{1}{7803--7821}.
\PrintBackRefs{\CurrentBib}

\bibitem [\protect \citeauthoryear {%
Schubert%
, Gula%
, Greatbatch%
, Baschek%
\BCBL {}\ \BBA {} Biastoch%
}{%
Schubert%
\ \protect \BOthers {.}}{%
{\protect \APACyear {2020}}%
}]{%
schubert2020submesoscale}
\APACinsertmetastar {%
schubert2020submesoscale}%
\begin{APACrefauthors}%
Schubert, R.%
, Gula, J.%
, Greatbatch, R\BPBI J.%
, Baschek, B.%
\BCBL {}\ \BBA {} Biastoch, A.%
\end{APACrefauthors}%
\unskip\
\newblock
\APACrefYearMonthDay{2020}{}{}.
\newblock
{\BBOQ}\APACrefatitle {The submesoscale kinetic energy cascade: Mesoscale
  absorption of submesoscale mixed layer eddies and frontal downscale fluxes}
  {The submesoscale kinetic energy cascade: Mesoscale absorption of
  submesoscale mixed layer eddies and frontal downscale fluxes}.{\BBCQ}
\newblock
\APACjournalVolNumPages{Journal of Physical Oceanography}{50}{9}{2573--2589}.
\PrintBackRefs{\CurrentBib}

\bibitem [\protect \citeauthoryear {%
Scott%
\ \BBA {} Arbic%
}{%
Scott%
\ \BBA {} Arbic%
}{%
{\protect \APACyear {2007}}%
}]{%
ScottArbic07}
\APACinsertmetastar {%
ScottArbic07}%
\begin{APACrefauthors}%
Scott, R\BPBI B.%
\BCBT {}\ \BBA {} Arbic, B\BPBI K.%
\end{APACrefauthors}%
\unskip\
\newblock
\APACrefYearMonthDay{2007}{{\APACmonth{03}}}{}.
\newblock
{\BBOQ}\APACrefatitle {{Spectral Energy Fluxes in Geostrophic Turbulence:
  Implications for Ocean Energetics}} {{Spectral Energy Fluxes in Geostrophic
  Turbulence: Implications for Ocean Energetics}}.{\BBCQ}
\newblock
\APACjournalVolNumPages{Journal of Physical Oceanography}{37}{3}{673--688}.
\PrintBackRefs{\CurrentBib}

\bibitem [\protect \citeauthoryear {%
Scott%
\ \BBA {} Wang%
}{%
Scott%
\ \BBA {} Wang%
}{%
{\protect \APACyear {2005}}%
}]{%
ScottWang05}
\APACinsertmetastar {%
ScottWang05}%
\begin{APACrefauthors}%
Scott, R\BPBI B.%
\BCBT {}\ \BBA {} Wang, F.%
\end{APACrefauthors}%
\unskip\
\newblock
\APACrefYearMonthDay{2005}{}{}.
\newblock
{\BBOQ}\APACrefatitle {{Direct Evidence of an Oceanic Inverse Kinetic Energy
  Cascade from Satellite Altimetry}} {{Direct Evidence of an Oceanic Inverse
  Kinetic Energy Cascade from Satellite Altimetry}}.{\BBCQ}
\newblock
\APACjournalVolNumPages{Journal of Physical Oceanography}{35}{}{1650}.
\PrintBackRefs{\CurrentBib}

\bibitem [\protect \citeauthoryear {%
{Sogge}%
}{%
{Sogge}%
}{%
{\protect \APACyear {2008}}%
}]{%
Sogge}
\APACinsertmetastar {%
Sogge}%
\begin{APACrefauthors}%
{Sogge}, C\BPBI D.%
\end{APACrefauthors}%
\unskip\
\newblock
\APACrefYear{2008}.
\newblock
\APACrefbtitle {{Fourier Integrals in Classical Analysis.}} {{Fourier Integrals
  in Classical Analysis.}}
\newblock
\APACaddressPublisher{}{Cambridge University Press, New York}.
\PrintBackRefs{\CurrentBib}

\bibitem [\protect \citeauthoryear {%
Srinivasan%
, McWilliams%
, Molemaker%
\BCBL {}\ \BBA {} Barkan%
}{%
Srinivasan%
\ \protect \BOthers {.}}{%
{\protect \APACyear {2019}}%
}]{%
srinivasan2019submesoscale}
\APACinsertmetastar {%
srinivasan2019submesoscale}%
\begin{APACrefauthors}%
Srinivasan, K.%
, McWilliams, J\BPBI C.%
, Molemaker, M\BPBI J.%
\BCBL {}\ \BBA {} Barkan, R.%
\end{APACrefauthors}%
\unskip\
\newblock
\APACrefYearMonthDay{2019}{}{}.
\newblock
{\BBOQ}\APACrefatitle {Submesoscale vortical wakes in the lee of topography}
  {Submesoscale vortical wakes in the lee of topography}.{\BBCQ}
\newblock
\APACjournalVolNumPages{Journal of Physical Oceanography}{49}{7}{1949--1971}.
\PrintBackRefs{\CurrentBib}

\bibitem [\protect \citeauthoryear {%
Stanley%
, Bachman%
\BCBL {}\ \BBA {} Grooms%
}{%
Stanley%
\ \protect \BOthers {.}}{%
{\protect \APACyear {2020}}%
}]{%
stanley2020vertical}
\APACinsertmetastar {%
stanley2020vertical}%
\begin{APACrefauthors}%
Stanley, Z.%
, Bachman, S.%
\BCBL {}\ \BBA {} Grooms, I.%
\end{APACrefauthors}%
\unskip\
\newblock
\APACrefYearMonthDay{2020}{}{}.
\newblock
{\BBOQ}\APACrefatitle {Vertical structure of ocean mesoscale eddies with
  implications for parameterizations of tracer transport} {Vertical structure
  of ocean mesoscale eddies with implications for parameterizations of tracer
  transport}.{\BBCQ}
\newblock
\APACjournalVolNumPages{Journal of Advances in Modeling Earth
  Systems}{12}{10}{e2020MS002151}.
\PrintBackRefs{\CurrentBib}

\bibitem [\protect \citeauthoryear {%
{Stein}%
\ \BBA {} {Weiss}%
}{%
{Stein}%
\ \BBA {} {Weiss}%
}{%
{\protect \APACyear {1971}}%
}]{%
SteinWeiss}
\APACinsertmetastar {%
SteinWeiss}%
\begin{APACrefauthors}%
{Stein}, E\BPBI M.%
\BCBT {}\ \BBA {} {Weiss}, G.%
\end{APACrefauthors}%
\unskip\
\newblock
\APACrefYear{1971}.
\newblock
\APACrefbtitle {{Introduction to Fourier Analysis on Euclidean Spaces.}}
  {{Introduction to Fourier Analysis on Euclidean Spaces.}}
\newblock
\APACaddressPublisher{}{Princeton University Press, Princeton, New Jersey}.
\PrintBackRefs{\CurrentBib}

\bibitem [\protect \citeauthoryear {%
Storer%
, Buzzicotti%
, Khatri%
, Griffies%
\BCBL {}\ \BBA {} Aluie%
}{%
Storer%
\ \protect \BOthers {.}}{%
{\protect \APACyear {2022}}%
}]{%
Storer2022}
\APACinsertmetastar {%
Storer2022}%
\begin{APACrefauthors}%
Storer, B\BPBI A.%
, Buzzicotti, M.%
, Khatri, H.%
, Griffies, S\BPBI M.%
\BCBL {}\ \BBA {} Aluie, H.%
\end{APACrefauthors}%
\unskip\
\newblock
\APACrefYearMonthDay{2022}{sep}{}.
\newblock
{\BBOQ}\APACrefatitle {{Global energy spectrum of the general oceanic
  circulation}} {{Global energy spectrum of the general oceanic
  circulation}}.{\BBCQ}
\newblock
\APACjournalVolNumPages{Nature Communications}{13}{1}{5314}.
\newblock
\begin{APACrefDOI} \doi{10.1038/s41467-022-33031-3} \end{APACrefDOI}
\PrintBackRefs{\CurrentBib}

\bibitem [\protect \citeauthoryear {%
Thomson%
\ \BBA {} Emery%
}{%
Thomson%
\ \BBA {} Emery%
}{%
{\protect \APACyear {2001}}%
}]{%
ThomsonEmery01}
\APACinsertmetastar {%
ThomsonEmery01}%
\begin{APACrefauthors}%
Thomson, R\BPBI E.%
\BCBT {}\ \BBA {} Emery, W\BPBI J.%
\end{APACrefauthors}%
\unskip\
\newblock
\APACrefYear{2001}.
\newblock
\APACrefbtitle {{Data Analysis Methods in Physical Oceanography}} {{Data
  Analysis Methods in Physical Oceanography}}\ (\PrintOrdinal{3rd}\ \BEd).
\newblock
\APACaddressPublisher{}{Elsevier Science}.
\PrintBackRefs{\CurrentBib}

\bibitem [\protect \citeauthoryear {%
Torres%
\ \protect \BOthers {.}}{%
Torres%
\ \protect \BOthers {.}}{%
{\protect \APACyear {2018}}%
}]{%
Torresetal2018jgr}
\APACinsertmetastar {%
Torresetal2018jgr}%
\begin{APACrefauthors}%
Torres, H\BPBI S.%
, Klein, P.%
, Menemenlis, D.%
, Qiu, B.%
, Su, Z.%
, Wang, J.%
\BDBL {}Fu, L\BHBI L.%
\end{APACrefauthors}%
\unskip\
\newblock
\APACrefYearMonthDay{2018}{{\APACmonth{11}}}{}.
\newblock
{\BBOQ}\APACrefatitle {{Partitioning Ocean Motions Into Balanced Motions and
  Internal Gravity Waves: A Modeling Study in Anticipation of Future Space
  Missions}} {{Partitioning Ocean Motions Into Balanced Motions and Internal
  Gravity Waves: A Modeling Study in Anticipation of Future Space
  Missions}}.{\BBCQ}
\newblock
\APACjournalVolNumPages{Journal of Geophysical
  Research-Oceans}{123}{11}{8084--8105}.
\PrintBackRefs{\CurrentBib}

\bibitem [\protect \citeauthoryear {%
Trenberth%
}{%
Trenberth%
}{%
{\protect \APACyear {1975}}%
}]{%
Trenberth1975}
\APACinsertmetastar {%
Trenberth1975}%
\begin{APACrefauthors}%
Trenberth, K\BPBI E.%
\end{APACrefauthors}%
\unskip\
\newblock
\APACrefYearMonthDay{1975}{}{}.
\newblock
{\BBOQ}\APACrefatitle {A quasi-biennial standing wave in the Southern
  Hemisphere and interrelations with sea surface temperature} {A quasi-biennial
  standing wave in the southern hemisphere and interrelations with sea surface
  temperature}.{\BBCQ}
\newblock
\APACjournalVolNumPages{Quarterly Journal of the Royal Meteorological
  Society}{101}{427}{55--74}.
\PrintBackRefs{\CurrentBib}

\bibitem [\protect \citeauthoryear {%
Uchida%
, Jamet%
, Poje%
\BCBL {}\ \BBA {} Dewar%
}{%
Uchida%
\ \protect \BOthers {.}}{%
{\protect \APACyear {2021}}%
}]{%
Uchidaetal2021}
\APACinsertmetastar {%
Uchidaetal2021}%
\begin{APACrefauthors}%
Uchida, T.%
, Jamet, Q.%
, Poje, A.%
\BCBL {}\ \BBA {} Dewar, W\BPBI K.%
\end{APACrefauthors}%
\unskip\
\newblock
\APACrefYearMonthDay{2021}{}{}.
\newblock
{\BBOQ}\APACrefatitle {An ensemble-based eddy and spectral analysis, with
  application to the {Gulf Stream}} {An ensemble-based eddy and spectral
  analysis, with application to the {Gulf Stream}}.{\BBCQ}
\newblock
\APACjournalVolNumPages{Journal of Advances in Modeling Earth
  Systems}{}{}{e2021MS002692}.
\PrintBackRefs{\CurrentBib}

\bibitem [\protect \citeauthoryear {%
{Vallis}%
}{%
{Vallis}%
}{%
{\protect \APACyear {2017}}%
}]{%
Vallis17}
\APACinsertmetastar {%
Vallis17}%
\begin{APACrefauthors}%
{Vallis}, G\BPBI K.%
\end{APACrefauthors}%
\unskip\
\newblock
\APACrefYear{2017}.
\newblock
\APACrefbtitle {{Atmospheric and Oceanic Fluid Dynamics}} {{Atmospheric and
  Oceanic Fluid Dynamics}}.
\newblock
\APACaddressPublisher{}{{Cambridge University Press}}.
\newblock
\APACrefnote{2nd edition}
\PrintBackRefs{\CurrentBib}

\bibitem [\protect \citeauthoryear {%
Vreman%
, Geurts%
\BCBL {}\ \BBA {} Kuerten%
}{%
Vreman%
\ \protect \BOthers {.}}{%
{\protect \APACyear {1994}}%
}]{%
vreman1994realizability}
\APACinsertmetastar {%
vreman1994realizability}%
\begin{APACrefauthors}%
Vreman, B.%
, Geurts, B.%
\BCBL {}\ \BBA {} Kuerten, H.%
\end{APACrefauthors}%
\unskip\
\newblock
\APACrefYearMonthDay{1994}{}{}.
\newblock
{\BBOQ}\APACrefatitle {Realizability conditions for the turbulent stress tensor
  in large-eddy simulation} {Realizability conditions for the turbulent stress
  tensor in large-eddy simulation}.{\BBCQ}
\newblock
\APACjournalVolNumPages{Journal of Fluid Mechanics}{278}{}{351--362}.
\PrintBackRefs{\CurrentBib}

\bibitem [\protect \citeauthoryear {%
Wieczorek%
\ \BBA {} Meschede%
}{%
Wieczorek%
\ \BBA {} Meschede%
}{%
{\protect \APACyear {2018}}%
}]{%
Wieczorek2018}
\APACinsertmetastar {%
Wieczorek2018}%
\begin{APACrefauthors}%
Wieczorek, M\BPBI A.%
\BCBT {}\ \BBA {} Meschede, M.%
\end{APACrefauthors}%
\unskip\
\newblock
\APACrefYearMonthDay{2018}{aug}{}.
\newblock
{\BBOQ}\APACrefatitle {{SHTools: Tools for Working with Spherical Harmonics}}
  {{SHTools: Tools for Working with Spherical Harmonics}}.{\BBCQ}
\newblock
\APACjournalVolNumPages{Geochemistry, Geophysics,
  Geosystems}{19}{8}{2574--2592}.
\newblock
\begin{APACrefDOI} \doi{10.1029/2018GC007529} \end{APACrefDOI}
\PrintBackRefs{\CurrentBib}

\bibitem [\protect \citeauthoryear {%
Wunsch%
}{%
Wunsch%
}{%
{\protect \APACyear {1991}}%
}]{%
wunsch1991global}
\APACinsertmetastar {%
wunsch1991global}%
\begin{APACrefauthors}%
Wunsch, C.%
\end{APACrefauthors}%
\unskip\
\newblock
\APACrefYearMonthDay{1991}{}{}.
\newblock
{\BBOQ}\APACrefatitle {Global-scale sea surface variability from combined
  altimetric and tide gauge measurements} {Global-scale sea surface variability
  from combined altimetric and tide gauge measurements}.{\BBCQ}
\newblock
\APACjournalVolNumPages{Journal of Geophysical Research:
  Oceans}{96}{C8}{15053--15082}.
\PrintBackRefs{\CurrentBib}

\bibitem [\protect \citeauthoryear {%
Wunsch%
}{%
Wunsch%
}{%
{\protect \APACyear {2007}}%
}]{%
wunsch2007past}
\APACinsertmetastar {%
wunsch2007past}%
\begin{APACrefauthors}%
Wunsch, C.%
\end{APACrefauthors}%
\unskip\
\newblock
\APACrefYearMonthDay{2007}{}{}.
\newblock
{\BBOQ}\APACrefatitle {The past and future ocean circulation from a
  contemporary perspective} {The past and future ocean circulation from a
  contemporary perspective}.{\BBCQ}
\newblock
\APACjournalVolNumPages{Geophysical Monograph-American Geophysical
  Union}{173}{}{53}.
\PrintBackRefs{\CurrentBib}

\bibitem [\protect \citeauthoryear {%
Wunsch%
\ \BBA {} Stammer%
}{%
Wunsch%
\ \BBA {} Stammer%
}{%
{\protect \APACyear {1995}}%
}]{%
wunsch1995global}
\APACinsertmetastar {%
wunsch1995global}%
\begin{APACrefauthors}%
Wunsch, C.%
\BCBT {}\ \BBA {} Stammer, D.%
\end{APACrefauthors}%
\unskip\
\newblock
\APACrefYearMonthDay{1995}{}{}.
\newblock
{\BBOQ}\APACrefatitle {The global frequency-wavenumber spectrum of oceanic
  variability estimated from TOPEX/POSEIDON altimetric measurements} {The
  global frequency-wavenumber spectrum of oceanic variability estimated from
  topex/poseidon altimetric measurements}.{\BBCQ}
\newblock
\APACjournalVolNumPages{Journal of Geophysical Research:
  Oceans}{100}{C12}{24895--24910}.
\PrintBackRefs{\CurrentBib}

\bibitem [\protect \citeauthoryear {%
Youngs%
, Thompson%
, Lazar%
\BCBL {}\ \BBA {} Richards%
}{%
Youngs%
\ \protect \BOthers {.}}{%
{\protect \APACyear {2017}}%
}]{%
Youngsetal2017}
\APACinsertmetastar {%
Youngsetal2017}%
\begin{APACrefauthors}%
Youngs, M\BPBI K.%
, Thompson, A\BPBI F.%
, Lazar, A.%
\BCBL {}\ \BBA {} Richards, K\BPBI J.%
\end{APACrefauthors}%
\unskip\
\newblock
\APACrefYearMonthDay{2017}{{\APACmonth{06}}}{}.
\newblock
{\BBOQ}\APACrefatitle {{ACC Meanders, Energy Transfer, and Mixed
  Barotropic--Baroclinic Instability}} {{ACC Meanders, Energy Transfer, and
  Mixed Barotropic--Baroclinic Instability}}.{\BBCQ}
\newblock
\APACjournalVolNumPages{Journal of Physical Oceanography}{47}{6}{1291--1305}.
\PrintBackRefs{\CurrentBib}

\bibitem [\protect \citeauthoryear {%
Zanna%
, Porta~Mana%
, Anstey%
, David%
\BCBL {}\ \BBA {} Bolton%
}{%
Zanna%
\ \protect \BOthers {.}}{%
{\protect \APACyear {2017}}%
}]{%
Zannaetal17}
\APACinsertmetastar {%
Zannaetal17}%
\begin{APACrefauthors}%
Zanna, L.%
, Porta~Mana, P.%
, Anstey, J.%
, David, T.%
\BCBL {}\ \BBA {} Bolton, T.%
\end{APACrefauthors}%
\unskip\
\newblock
\APACrefYearMonthDay{2017}{{\APACmonth{03}}}{}.
\newblock
{\BBOQ}\APACrefatitle {{Scale-aware deterministic and stochastic
  parametrizations of eddy-mean flow interaction}} {{Scale-aware deterministic
  and stochastic parametrizations of eddy-mean flow interaction}}.{\BBCQ}
\newblock
\APACjournalVolNumPages{Ocean Modelling}{111}{}{66--80}.
\PrintBackRefs{\CurrentBib}

\bibitem [\protect \citeauthoryear {%
Zhao%
\ \BBA {} Aluie%
}{%
Zhao%
\ \BBA {} Aluie%
}{%
{\protect \APACyear {2018}}%
}]{%
ZhaoAluie18}
\APACinsertmetastar {%
ZhaoAluie18}%
\begin{APACrefauthors}%
Zhao, D.%
\BCBT {}\ \BBA {} Aluie, H.%
\end{APACrefauthors}%
\unskip\
\newblock
\APACrefYearMonthDay{2018}{{\APACmonth{05}}}{}.
\newblock
{\BBOQ}\APACrefatitle {{Inviscid criterion for decomposing scales }} {{Inviscid
  criterion for decomposing scales }}.{\BBCQ}
\newblock
\APACjournalVolNumPages{Physical Review Fluids}{3}{5}{301}.
\PrintBackRefs{\CurrentBib}

\end{thebibliography}

\end{document}